\documentclass[tighten,twocolumn]{aastex63}
\accepted{\today}
\submitjournal{ApJ}

\shorttitle{PSR J0212+5321 Discovery}
\shortauthors{Perez et al.}

\graphicspath{{./}{figures/}}
\usepackage{glossaries}

\begin{document}

\title{Green Bank Telescope Discovery of the Redback Binary Millisecond Pulsar PSR J0212+5321}

\newcommand{\BLI}{Breakthrough Listen Initiative\xspace}
\newcommand{\BL}{Breakthrough Listen\xspace}

\newcommand{\GBT}{\textit{Green Bank Telescope}\xspace}

\newcommand{\subtot}[1]{{\textcolor{Cerulean}{#1}}}
\newcommand{\subsubtot}[1]{\textit{\textcolor{OliveGreen}{#1}}}
\newcommand{\tot}[1]{{\textcolor{NavyBlue}{#1}}}
\newcommand{\ie}{i.\,e.\,}
\newcommand{\eg}{e.\,g.\,}

\newcommand{\UCB}{Department of Astronomy,  University of California Berkeley, Berkeley CA 94720}
\newcommand{\SETI}{SETI Institute, Mountain View, California}

\correspondingauthor{Karen I.~Perez}
\email{karen.i.perez@columbia.edu}

\author[0000-0002-6341-4548]{Karen I.~Perez}
\affiliation{Department of Astronomy, Columbia University, 550 West 120th Street, New York, NY 10027, USA}

\author[0000-0002-9870-2742]{Slavko Bogdanov}
\affiliation{Columbia Astrophysics Laboratory, Columbia University, 550 West 120th Street, New York, NY 10027, USA}

\author[0000-0003-4814-2377]{Jules P.~Halpern}
\affiliation{Department of Astronomy, Columbia University, 550 West 120th Street, New York, NY 10027, USA}
\affiliation{Columbia Astrophysics Laboratory, Columbia University, 550 West 120th Street, New York, NY 10027, USA}

\author[0000-0002-8604-106X]{Vishal Gajjar}
\affiliation{\UCB}

\begin{abstract}
We report the discovery of a 2.11\,ms binary millisecond pulsar during a targeted search of the redback optical candidate coincident with the $\gamma$-ray source 3FGL J0212.5+5320 using the Robert C. Byrd Green Bank Telescope (GBT) with the Breakthrough Listen backend at L-band.  Over a seven month period, five pointings were made near inferior conjunction of the pulsar in its 20.9\,hr orbit, resulting in two detections, lasting 12 and 42 minutes. The pulsar dispersion measure (DM) of 25.7~pc\,cm$^{-3}$ corresponds to a distance of 1.15~kpc in the NE2001 Galactic electron density model, consistent with the Gaia parallax distance of $1.16\pm0.03$~kpc for the companion star. We suspect the pulsar experiences wide-orbit eclipses, similar to other redbacks, as well as scintillation and DM delays caused by its interaction with its companion and surroundings. Although the pulsar was only detected over $\approx3.7\%$ of the orbit, its measured acceleration is consistent with published binary parameters from optical radial velocity spectroscopy and light-curve modeling of the companion star, and it provides a more precise mass ratio and a projected semi-major axis for the pulsar orbit.  We also obtained a refined optical photometric orbit ephemeris, and observed variability of the tidally distorted companion over 7 years.  A hard X-ray light curve from NuSTAR shows expected orbit-modulated emission from the intrabinary shock.  The pulsar parameters and photometric ephemeris greatly restrict the parameter space required to search for a coherent timing solution including pulsar spin-down rate, either using Fermi $\gamma$-rays, or further radio pulse detections.
 
\end{abstract}

\keywords{Pulsars; Millisecond Pulsars; Redbacks}

\section{Introduction} 
\label{sec:intro}
Compared to the population of ordinary pulsars, millisecond pulsars (MSPs) have much shorter spin periods ($P_{s}$ $\lesssim$ 30 ms), smaller spin-down rates ($\dot{P_s}$ $\sim$ $10^{-22}$ -- $10^{-17}$ s/s), weak magnetic fields (B$_{s}$ $\sim$ $10^{7}$ -- $10^{11}$), and long spin-down timescales of $\tau$ $\sim$ $10^{9}$ -- $10^{10}$ years \citep{Hui_2019}. Since their discovery by \citet{Backer_1982}, it has been commonly accepted that MSPs have been recycled by the transfer of matter and angular momentum from a companion star during an X-ray binary phase \citep{Alpar_1982}. This has been supported by the discovery of three transitional millisecond pulsars (tMSPs). They are characterized by instantaneous switches between two clearly distinguishable states: an accreting low-mass X-ray binary (LMXB) and a rotationally-powered MSP -- both of which last for several years in a single instance (\citealt{Archibald_2009, Pappito_2013, Bassa_2014, Stappers_2014}). This interaction is due to the balance between the outward pressure exerted by the pulsar wind on the mass lost by its companion, and the inward pull applied by the gravitational field of the pulsar (\citealt{BHATTACHARYA_1991, Papitto_2020}). While the more common MSP binaries have white dwarf companions and wide orbits ($P_b>1$~day), the interacting MSP systems have a low-mass stellar or sub-stellar companion, and orbits that have been circularized through tidal interactions, with $P_b<1$~day (\citealt{Fruchter_1988, Roberts_2013}). Black widows have degenerate companions with $M_{2}$ $<$ 0.05 $M_{\odot}$, while redbacks have hydrogen-rich, non-degenerate companions (Roche-lobe filling or nearly so) with 0.1 $M_{\odot}$ $\lesssim$ $M_{2}$ $\lesssim$ 1 $M_{\odot}$ \citep{Strader_2019}. Deriving their names from spider species which prey on their mates, pulsars in these binary systems ablate their companions and possibly destroy some completely to become isolated MSPs.

Redbacks are a growing class of binary MSPs. Originally discovered in globular clusters (\citealt{Camilo_2000,Edmonds_2002, dam01, Ransom_2005}), there are currently over two dozen such systems known in the field of the Galaxy, either confirmed or candidate (\citealt{Roberts_2013, Bogdanov2021}), demonstrating that they are not exclusive to globular clusters as initially thought. The discovery of tMSPs has been key to investigating the evolution of MSPs, as we work towards understanding whether they are solely an intermediate step before they end as radio pulsars devouring their companions, or are instead a road to a different evolutionary path. 

The radio eclipse properties of redback systems are highly variable, and they are strongly frequency dependent, with generally much longer eclipses at lower frequencies. In addition, at lower frequencies, random, brief eclipses are observed at all phases \citep{Archibald_2009}. Such eclipses and dispersion measure variations are attributed to the presence of gas flowing out from the companion due to irradiation by the pulsar wind.

Redbacks are detected with surprising frequency in high-energy $\gamma$-rays by the Fermi Large Area Telescope (LAT), and are identified as such by follow-up radio pulsar searches of Fermi unassociated $\gamma$-ray sources. A number of Fermi LAT sources have been found to contain optical and X-ray counterparts with properties virtually identical to those of confirmed redback MSPs, but no radio pulsations have been found so far. In many cases, redback MSPs are discovered in the radio only after several observations of the source. Multiple observations increase the chances that an intermittently detectable pulsar will be glimpsed despite the interstellar scintillation, large acceleration at an unfavorable binary phase, or eclipses caused by the material emanating from the companion star.

In this paper, we present the discovery of the redback binary MSP PSR J0212+5321, along with optical and X-ray observations of the 3FGL J0212.1+5320 system. In Section \ref{sect:summary} we summarize the $\gamma$-ray, optical, and X-ray properties indicating 3FGL J0212.1$+$5320 to be a redback MSP. In Section \ref{sect:obs_stratergy} we outline our observations and observing strategy with the Green Bank Telescope (GBT) at L-band, and detail our search, analysis, and discovery in Section \ref{sect:data}. Section \ref{sect:opt} describes our optical monitoring observations and refined optical ephemeris using the MDM Observatory and the Zwicky Transient Facility (ZTF). Section \ref{sect:xray} covers X-ray monitoring observations with the Nuclear Spectroscopic Telescope Array (NuSTAR), Swift, and Chandra. We discuss the implications of our findings and derived parameters in Section \ref{sect:discussion}, and Section \ref{sect:conclusion} lists our final conclusions. 

\section{History of 3FGL J0212.1+5320}
\label{sect:summary}
The high-energy $\gamma$-ray source 3FGL J0212.1$+$5320 (presently cataloged as 4FGL J0212.1+5321) first appeared in the initial Fermi LAT source catalog \citep{Abdo_2010}. It was deemed to be a strong MSP candidate through the use of statistical and machine-learning techniques (\citealt{Saz_2016, Mirabal_2016}).  \citet{Linares_2017} found 3FGL J0212.1$+$5320 to have a low variability and a curved $\gamma$-ray spectral shape, both $\gamma$-ray properties of pulsars from the \textit{Fermi} LAT catalog \citep{Abdo_2013}. 

Follow-up photometry and optical spectroscopy by \citet{Li_2016} and \citet{Linares_2017} found a variable optical and X-ray counterpart to the source -- a nearby ($\approx 1.1$~kpc) binary system with a 20.9 hour period and an optically bright ($r'\approx14.3$~mag) tidally deformed main-sequence secondary star of spectral type $\approx$\,F6 and mass $\approx0.4\,M_{\odot}$, making it the brightest in the optical band out of all known redbacks (confirmed and candidates). \citet{Li_2016} determined an orbital period $P_b$ = 0.869575(4) days, which we further refine in this paper. \citet{Linares_2017} found that its optical lightcurve was similar to those observed in the confirmed redback MSPs PSR J1628$-$3205 ($P_b=5.0$\,hr; \citealt{Li_2014}) and PSR J2129$-$0429 ($P_b=15.2$\,hr; \citealt{Bellm_2016}).
Detailed modeling of the optical light curve by \citet{sha17}, including heating or starspot effects, put further constraints on the binary parameters such as mass ratio and inclination angle.

The X-ray-to-$\gamma$-ray flux ratio and $\gamma$-ray luminosity of $L_{\gamma} = 2.5 \times10^{33}$ erg\,s$^{-1}$ are also typical of MSPs. \textit{Chandra} X-ray observations revealed a power-law spectrum with $\Gamma\approx 1.3$ and luminosity of $L_{X}$ = 2.6 $\times$ $10^{32}$ erg\,s$^{-1}$, consistent with the presence of a synchrotron-emitting intrabinary shock and making it the brightest non-accreting redback candidate in the X-rays.

\section{Strategy and GBT Observations}
\label{sect:obs_stratergy}
We observed the optical/X-ray candidate for 3FGL J0212.1$+$5320 at L-band (central frequency of $\nu_c$=1501 MHz) on five dates between 2022 February 11 and 2022 September 5 with the GBT using the Breakthrough Listen Digital Backend (BLDB; \citealt{MacMahon_BLDR}) to conduct a deep search for radio pulsations and confirm its identification as a binary MSP. We opted for the L-band frequency range because it is advantageous to observe at the highest frequency feasible, as eclipses tend to last longer at lower frequencies. On the other hand, L-band is an adequate compromise for MSPs, which generally have relatively steep radio spectra indices, ranging from approximately $-3$ to $-1.5$ \citep{Frail_2016}.

Our observations covered 1.3--1.85 GHz, with an effective bandwidth of 550 MHz. We collected baseband voltages and converted them to high-time resolution total intensity filterbank products. The number of observing channels was 16,384, with a channel bandwidth of $91$ kHz, and a sampling time of $43\,\mu$s. A detailed summary of the BL reduction pipeline and discussion of various data products is provided by  \cite{Lebofsky_2019}.

A summary of all our on-source observations, including observation lengths and corresponding orbital phases, is given in Table~\ref{tab:radio_obs}.  We observed the binary at around pulsar inferior conjunction (orbital phase $\phi=0.75$), when the pulsar is between the companion and the observer, to minimize the chance of eclipses. We computed the orbital phase using a refined optical photometric ephemeris obtained at the MDM Observatory; this is further discussed in Section~\ref{sec:ephem}.  Observations were separated by at least a month to mitigate against possible changes in eclipse duration on timescales of weeks/months, as seen in observations of redbacks such as PSR J1048$+$2339 \citep{Deneva_2016}. Our detections are listed in Table \ref{tab:radio_PRESTO_params}, and will be discussed further below.

\begin{deluxetable*}{ccccccc}[t]
\tablecolumns{7}
\tablewidth{0pt}
\tablecaption{Log of GBT Observations of 3FGL J0212.1+5320\label{tab:radio_obs}}
\tablehead{
    \colhead{Obs} & \colhead{Date} & \colhead{Start/Stop Time} & \colhead{Start/Stop Time} & \colhead{Orbital Phase\tablenotemark{a}} & \colhead{Total Exposure} & \colhead{Sub-integration\tablenotemark{b}} \\
    & (UT) & (UTC) & (BMJD TDB) & ($\phi$) & (min) & (min)}
    \startdata
    \hline
    01 & 2022--02--11 & 04:22:39 -- 04:30:26 & 59621.18309 -- 59621.18850 & 0.769 -- 0.775 & 7.8 & 7.8 \\
    02 & 2022--03--10/11 & 23:48:38 -- 00:30:02 & 59648.99068 -- 59649.01942 &  0.748 -- 0.781 & 41.4 & 5.175 \\
    03 & 2022--04--19 & 23:00:38 -- 23:15:23 & 59688.95527 -- 59688.96551& 0.706 -- 0.718 & 14.7 & 3.675 \\
    04 & 2022--05--24 & 18:27:01 -- 19:15:20 & 59723.76500 -- 59723.79856 & 0.737 -- 0.776 & 48.3 & 6.0375 \\ 
    05 & 2022--09--05 & 04:59:39 -- 05:45:39 & 59827.21055 -- 59827.24250 & 0.698 -- 0.735 & 46.01 & 5.75 \\
    \hline
    \enddata
    \tablenotetext{a}{Predicted from the optical photometric ephemeris (Section \ref{sec:ephem}).}
    \tablenotetext{b}{The length the observation was equally split into to thoroughly search for the pulsar.}
\end{deluxetable*}

\begin{deluxetable*}{cccccc}[t]
\tablecolumns{6}
\tablewidth{0pt}
\tablecaption{PRESTO Parameters of PSR J0212+5321 Detections with the GBT \label{tab:radio_PRESTO_params}}
\tablehead{
    \colhead{Obs} & \colhead{DM} & \colhead{$T_{\rm epoch}$} & \colhead{$P_s$\tablenotemark{a}} & \colhead{$\dot P_s$\tablenotemark{a}} & \colhead{Significance\tablenotemark{b}} \\ 
    & \colhead{(pc\,cm$^{-3}$)} & \colhead{(BMJD TDB)} & \colhead{(ms)} & \colhead{(s s$^{-1}$)} & \colhead{($\sigma$) }} 
    \startdata 
    \hline
    03 & 25.68 & 59688.95527 & 2.11459814(16) & 2.99(14) $\times$ $10^{-11}$ & 14.6 \\ 
    05 & 25.72 & 59827.21055 & 2.114580210(38) & 2.987(11) $\times$ $10^{-11}$ & 23.9 \\ 
    \enddata
    \tablenotetext{a}{Apparent spin period and period derivative at $T_{\rm epoch}$.}
    \tablenotetext{b}{Equivalent Gaussian significance, based on the probability of seeing a noise value with the same incoherently summed power \citep{Ransom_2002}.}

\end{deluxetable*}

\begin{deluxetable*}{ll}
\tablecolumns{2}
\tablewidth{0pt}
\tablecaption{Parameters of PSR J0212+5321\label{tab:pulsar_params}}
\tablehead{
    \colhead{Parameter} & \colhead{Value}} 
    \startdata
    \hline 
    R.A. (Gaia-CRF3)\tablenotemark{a}, $\alpha$ & $02^{\rm h}12^{\rm m}10^{\rm s}\!.47274$ \\
    Decl. (Gaia-CRF3)\tablenotemark{a}, $\delta$ & 53\degr 21\arcmin 38\farcs8110 \\
    Epoch of position\tablenotemark{a} & 2016.0 \\
    Proper motion\tablenotemark{a} in R.A., $\mu_{\rm \alpha}$\,cos\,$\delta$\, (mas yr$^{-1}$) & 
    $-2.627(23)$ \\
    Proper motion\tablenotemark{a} in decl., $\mu_{\rm \delta}$ (mas yr$^{-1}$) & 
    +2.044(21)\\
    Parallax distance\tablenotemark{a}, $d$ (kpc) & 1.16(03) \\
    Epoch of pulsar ascending node at $\phi = 0$\tablenotemark{b}, $T_0$ (BMJD TDB) & 58487.45679(86) \\
    Orbital period\tablenotemark{b}, $P_{b}$ (days) & 0.8695759(11) \\
    Intrinsic spin period\tablenotemark{c}, $P_{s,0}$ (ms) & 2.1147(1)\\
    Mass ratio\tablenotemark{d}, $q=M_2/M_1$ & $0.2469 \pm 0.0067 \pm 0.010$ \\
    Projected semi-major axis\tablenotemark{d}, $(a_1/c)\,{\rm sin}\,i$ (lt-s) & $2.132 \pm 0.008 \pm 0.09$ \\   
    Dispersion Measure\tablenotemark{e}, DM (pc\,cm$^{-3}$) & 25.7 \\
    DM distance based on NE2001, $d_{\rm{NE2001}}$ (kpc) & 1.15 \\
    DM distance based on YMW16, $d_{\rm{YMW16}}$ (kpc) & 1.27 \\
    1500 MHz flux density\tablenotemark{e}, $S_{1500}$ (mJy) & 0.04 \\
    1500 MHz radio luminosity\tablenotemark{e}, $L_{1500}$ = $S_{1500}$ $d^{2}$ (mJy\,kpc$^{2}$) & 0.05 \\
    Pulse width\tablenotemark{e} at 1500 MHz, $W_{50}$  (ms) & 0.176 \\
    \enddata
    \tablenotetext{a}{From Gaia DR3 \citep{Gaia_2022j}.}
    \tablenotetext{b}{From the MDM optical photometric ephemeris (Section~\ref{sec:ephem}).}
    \tablenotetext{c}{Derived in Section~\ref{sect:orbital_params}. Uncertainty is dominated by possible systematic error in orbital phase.}
    \tablenotetext{d}{Derived in Section~\ref{sect:orbital_params}. The second uncertainty represents possible systematic error in orbital phase.}
    \tablenotetext{e}{Measured from GBT Obs 05.}   
    
\end{deluxetable*}

\section{GBT Data Analysis and Results}
\label{sect:data}

\subsection{Pulse Search and Discovery}
\label{sect:search}
All observations were processed and searched for pulsars using standard \texttt{PRESTO}\footnote{Available for download from \url{https://github.com/scottransom/presto}.} routines \citep{Ransom_2001}. Radio frequency interference (RFI) masks were created for each observation using \texttt{rfifind}. We used the RFI mask with \texttt{prepsubband} to generate de-dispersed time series across 0--60 pc cm$^{-3}$, with a DM step of 0.1 pc cm$^{-3}$ without any restrictions on the spin period. When generating these time series, we fixed the pulsar position at the Gaia Data Release 3 (DR3) coordinates including proper motion (\citealt{Gaia_2016, Gaia_2022j}) as listed in Table~\ref{tab:pulsar_params}, and corrected the times to the solar system barycenter. We then Fourier transformed the time series with \texttt{realfft}. The resulting power spectra were summed up to the eighth harmonic and a Fourier-domain acceleration search was performed to search for periodic signals using \texttt{accelsearch} with z$_{\rm max}=200$, where z$_{\rm max}$ indicates the largest number of Fourier bins by which a pulse is allowed to smear across frequency. We do not conduct jerk searches (for changing acceleration), as J0212+5321 is not highly accelerated, and sensitivity would only improve for observation lengths $T_{\rm obs}\sim0.05-0.15\, P_b$ \citep{Andersen_2018}, corresponding to 1--3 hr long observations for this orbital period. Using the \texttt{sifting} script, candidates without duplicates or DM problems were folded with \texttt{prepfold} for visual inspection, where the number of sampling bins  was automatically selected based on the period. 

\begin{figure*}[!tbp]
  \centering
  \includegraphics[scale=0.75]{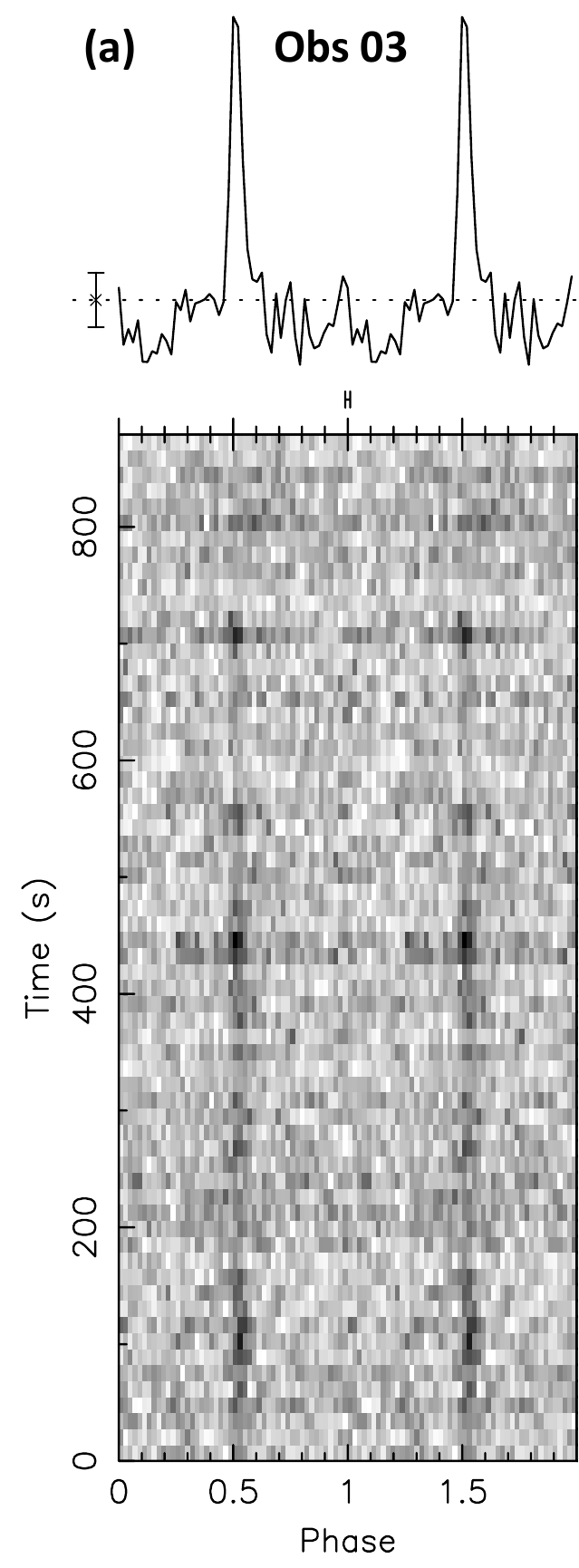}
  \label{fig:}
  \hspace*{0.5cm}
  \includegraphics[scale=0.75]{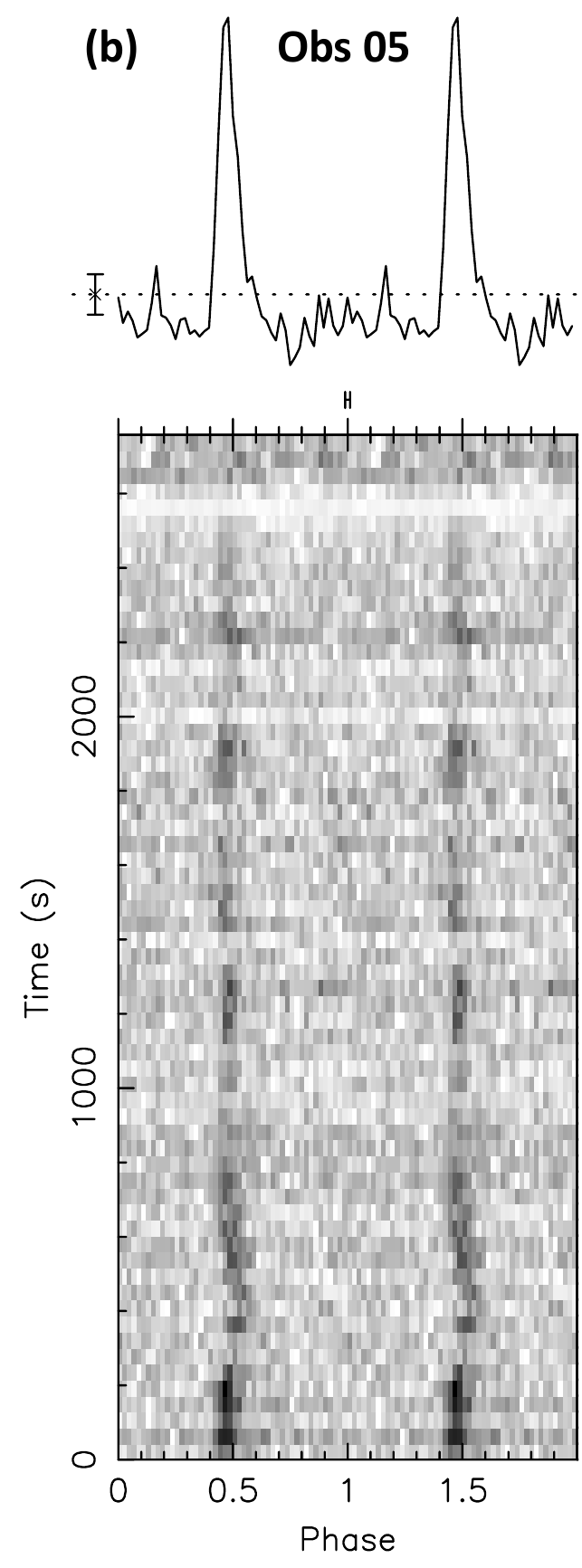}
  \caption{Time-resolved radio detections of PSR J0212+5321 taken at 1500 MHz with the GBT from (a) 2022 April 19 and (b) September 5, respectively. The length of the observations are 14.7 and 46 minutes, respectively. The folded pulse profiles are shown twice as a function of time and summed at the top.}
  \label{fig:radio_phase_vs_time}
\end{figure*}

\begin{figure*}[!tbp]
  \centering
  \includegraphics[scale=0.7]{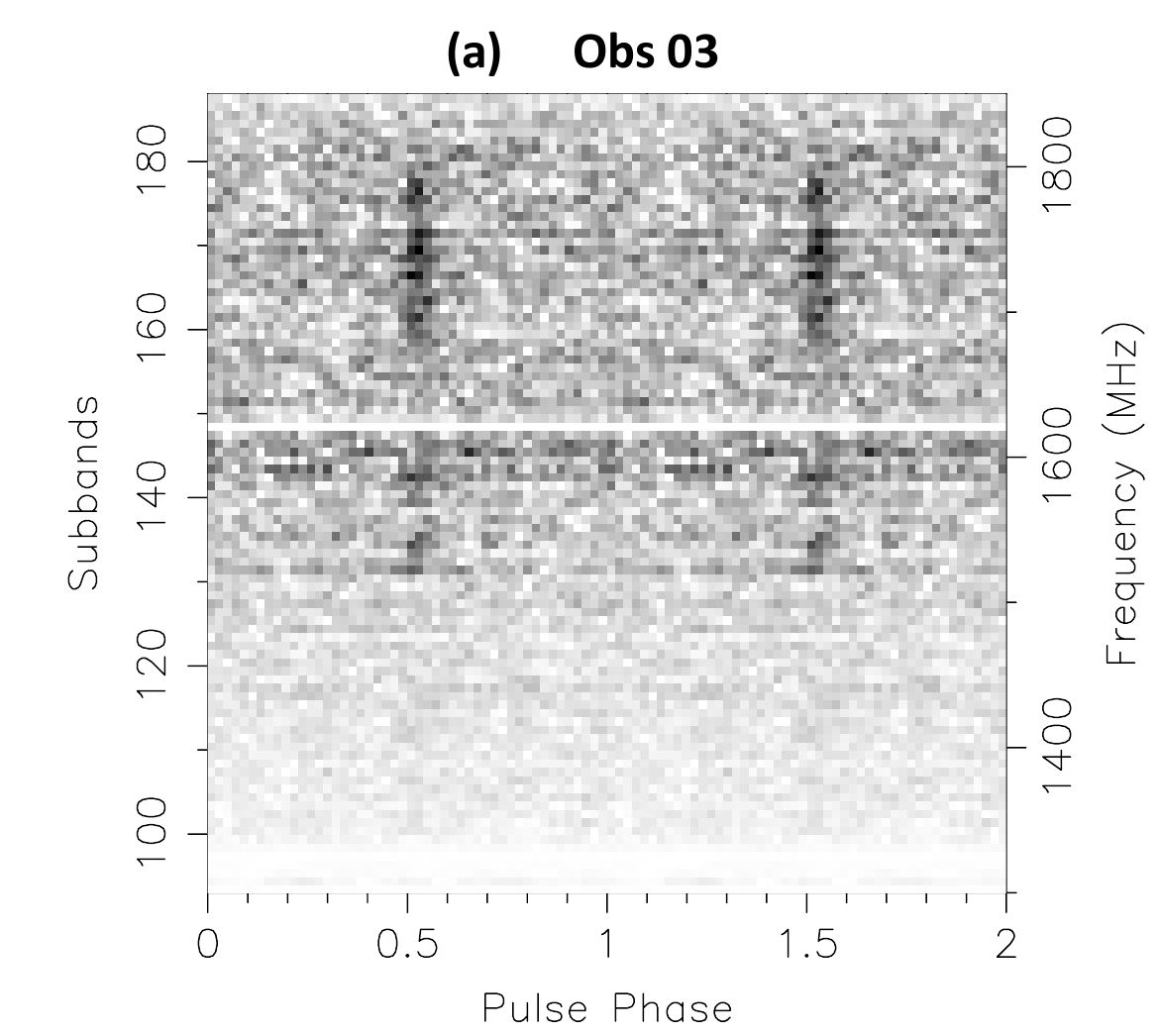}
  \hspace*{0.2cm}
  \includegraphics[scale=0.7]{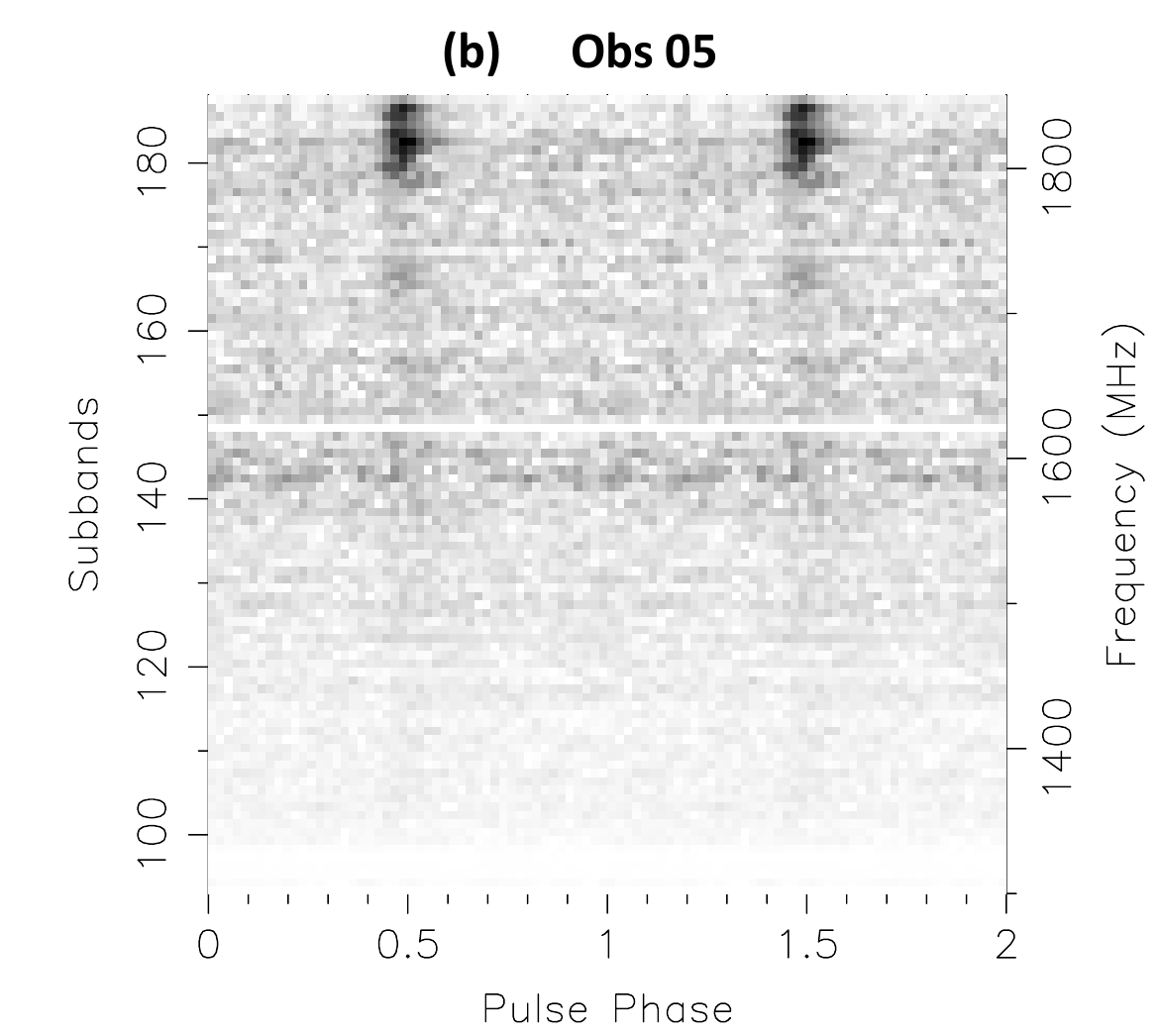}
  \caption{Sub-band plot for Obs 03 and 05, respectively, where 48 profile bins were used and each sub-band is 5.86~MHz wide for a total of 256 sub-bands. For data clarity, only sub-bands corresponding to frequencies 1.3--1.85 GHz (sub-bands 93--188) are shown.}
  \label{fig:radio_phase_vs_subband}
\end{figure*}

The main candidate was found in the full Obs 03 integration from 2022 April 19 at a period of 2.11 ms and DM of 25.7 pc\,cm$^{-3}$, using 48 sampling bins. Its detection parameters are listed in Table \ref{tab:radio_PRESTO_params}. Shown in Figure \ref{fig:radio_phase_vs_time}a, the signal was seen for the first $\approx720$ seconds, corresponding to orbital phases 0.706--0.716, just before pulsar inferior conjunction. Using the effective bandwidth, Figure \ref{fig:radio_phase_vs_subband}a shows the frequencies at which PSR J0212+5321 was detected. The pulsar dominates the top-half of the band from $\sim1500-1800$~MHz, indicating a low-frequency cutoff $\sim1500$~MHz. 

We folded the raw data for all other observations at the identified period and DM, at both the full- and sub-integrations of the observations (considering the possibility of brief eclipses), as noted in Table \ref{tab:radio_obs}. For observations longer than 10 minutes, the full observation was split into sub-integrations according to its length. Obs 02, 04, and 05 were divided into eight sub-integrations, while Obs 03 was divided into four. Given the Doppler effect on detection parameters, we additionally ensured a thorough search by folding each full observation and its sub-integrations at a range of spin periods and derivatives for a circular orbit, given their respective phases and accounting for systematic errors (see Section \ref{sect:orbital_params} for more details). We allowed for both a search in DM space, and a targeted search at the previously detected DM of 25.7 pc\,cm$^{-3}$.

While we did not detect the pulsar in Obs 01, 02, and 04, observations conducted on 2022 September 5 (Obs 05) resulted in a second detection of PSR J0212+5321, with similar detection parameters as listed in Table \ref{tab:radio_PRESTO_params}. As shown in Figure \ref{fig:radio_phase_vs_time}b, similar to Obs 03, the pulsar is visible through almost the entire session of Obs 05, $\approx42$~min out of 46~min, corresponding to orbital phases 0.698--0.732, with occasional dimming most likely due to intrabinary material surrounding the pulsar. Unlike Obs 03, Figure \ref{fig:radio_phase_vs_subband}b shows the pulsar much dimmer at mid- and lower-frequencies, peaking in intensity from $\sim$ 1790--1850 MHz. This leads us to believe the pulsar might be easily detectable at higher frequencies, e.g., S--band (2--4 GHz), or else the narrow-band excess could be a manifestation of interstellar scintillation. 

For Obs 05, we strategically chose to observe the pulsar at orbital phases closer to that of Obs 03, in case there is persistent visibility confined to phases preceding pulsar inferior conjunction. However, we cannot yet conclude that these two detections were strictly determined by orbital phase rather than random variability (see Section \ref{sect:eclipses} for a discussion). 

\subsection{Sensitivity and Flux Density}
A test pulsar, PSR B0329+54, was observed for 5 minutes prior to every observing session to verify our equipment sensitivity. We recovered it during all five of our observations with blind \texttt{PRESTO} searches at z$_{max}$=0.  We also detected harmonics of the slow 376 ms pulsar, PSR J0212+5222 (\citealt{Barr_2013, Lynch_2013}), at its known DM of 38.2 pc cm$^{-3}$ in the side-lobes of the telescope for all five observations. 

While our observations were not flux-calibrated, here we estimate our pulsar sensitivity using standard methods. Using \texttt{dspsr}'s \texttt{pdmp} to optimize the S/N of the pulse profile, we compute the flux density for each of the two observations using the radiometer equation \citep{Lorimer_2004}
\begin{equation}
    \label{radiometer_eqn}
    S_{\rm 1500} = \frac{S/N\,T_{\rm sys}}{G \,\sqrt{n_{p}\,\Delta\nu\,T_{\rm obs}}} \, \sqrt{\frac{W_{50}}{P_s-W_{50}}},
\end{equation}
where $S/N$ is the signal-to-noise ratio of PSR J0212+5321, $T_{\rm sys}$ is the equivalent temperature of the system (receiver, spillover, and sky), $n_{p}$ is the number of polarizations, $G$ is the telescope gain, and $\Delta\nu$ is the effective bandwidth. $T_{\rm obs}$ is the integration time, and $W_{50}$ and $P_s$ are the absolute pulse width and apparent spin period, respectively. We use $n_{p}=2, G=2.0$~K~Jy$^{-1}, T_{\rm sys}=17.1$~K, and $\Delta\nu=550$~MHz. We use an S/N of 16.66 and 27.29, and a $W_{50}$ of 0.132 ms and 0.176 ms, for Obs 03 and 05, respectively. We derive a flux density, $S_{\rm 1500}$ of 0.037 mJy and 0.040 mJy for Obs 03 and 05, respectively. Using the Gaia-derived distance, this corresponds to a radio luminosity, $L_{\rm 1500}$ of 0.05 mJy\,kpc$^{2}$. We report the pulsar parameters measured with Obs 05 in Table \ref{tab:pulsar_params}.

\subsection{Distance}
Gaia's DR3 parallax for PSR J0212+5321 corresponds to a distance of 1.16 $\pm$ 0.03 kpc. We also estimate the distance using two Galactic electron density distribution models from \citet{Price_2021}'s \texttt{PyGEDM}. Using the DM of 25.7 pc cm$^{-3}$ at Galactic coordinates of $l=134\degr\!.92$, $b=-7\degr\!.62$, the NE2001 (\citealt{Cordes_2002_NE2001, Cordes_2003_NE2001model}) model predicts $d_{\rm NE2001}$ = 1.15 kpc, while the YMW16 \citep{Yao_2017} model predicts $d_{\rm YMW16}$ = 1.27 kpc. The NE2001 model is most consistent with Gaia.

\subsection{Timing Anomaly}
\label{sect:timing_anomaly}
During Obs 05, there is a timing anomaly near the beginning (Figure \ref{fig:radio_phase_vs_time}b) lasting for $\approx400$~s during which the pulse is delayed by $\approx0.1\,P_s$.  The cause could be transient extra plasma in the system with electron column $\delta{\rm DM} \approx 0.13$~pc~cm$^{-3}$ (using an average frequency of 1600~MHz).
If originating in a spherical clump transiting the line of sight with the relative velocity of the stars ($\approx275$~km~s$^{-1}$), it would have an electron density of $\approx3.7\times10^7$~cm$^{-3}$ (see Section \ref{sect:eclipses} for further discussion).

\section{Optical Time-Series Monitoring\label{sect:opt}}

\begin{figure*}
\vspace{-1.4in}
\centerline{
\includegraphics[angle=0.,width=1.08\linewidth]{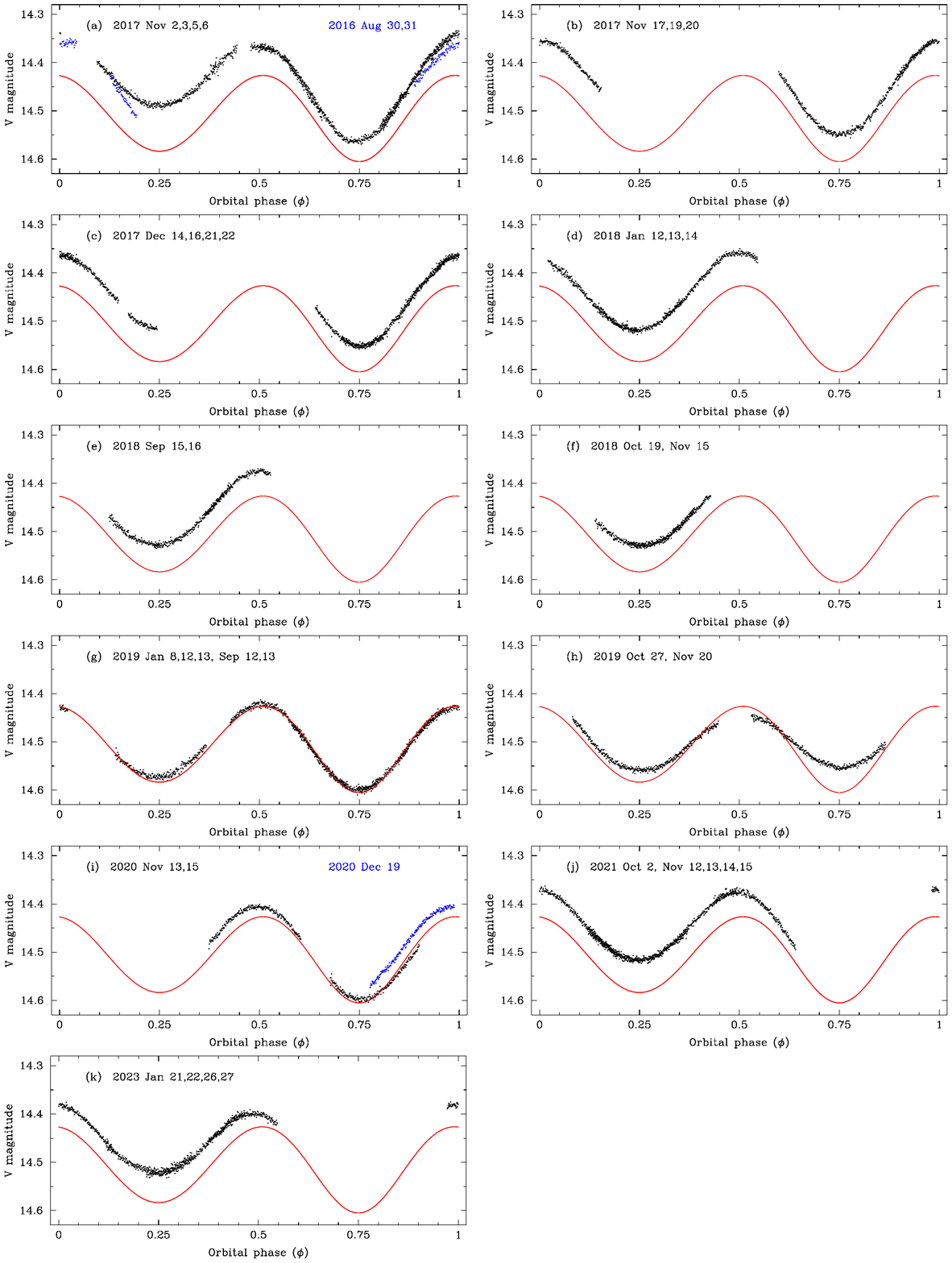}
}
\vspace{-0.5in}
  \caption{$V$-band time-series photometry of 3FGL J0212.1+5320 from
    the MDM 1.3~m, folded according to the orbital ephemeris calculated
    from these data (see Section~\ref{sec:ephem}).  Dates are given in each
    panel.  The red curve is a pure ellipsoidal model \citep{mor93,gom21}
    using the fitted orbital parameters (mass ratio, inclination angle,
    and Roche-lobe filling factor) from \citet{sha17}.  It is normalized
    to the data in panel (g), but is otherwise not a fit to the data.
}
\label{fig:opt}
\end{figure*}

We collected optical time-series of 3FGL J0212.1+5320 from 2016 to 2023 in order to monitor for variability, and to extend its orbital ephemeris for use in conjunction with the pulsar search.  The data were obtained with the MDM Observatory 1.3~m McGraw-Hill telescope in the $V$ band at 63~s cadence, and were reduced using differential photometry with respect to a nearby comparison star.   Figure~\ref{fig:opt} shows the light curves from the 39 nights, in time order from panels (a) through (k), folded on the orbital ephemeris (see Section~\ref{sec:ephem}).

\begin{figure*}
\vspace{-2.in}
\centerline{
\includegraphics[angle=0.,width=1.08\linewidth]{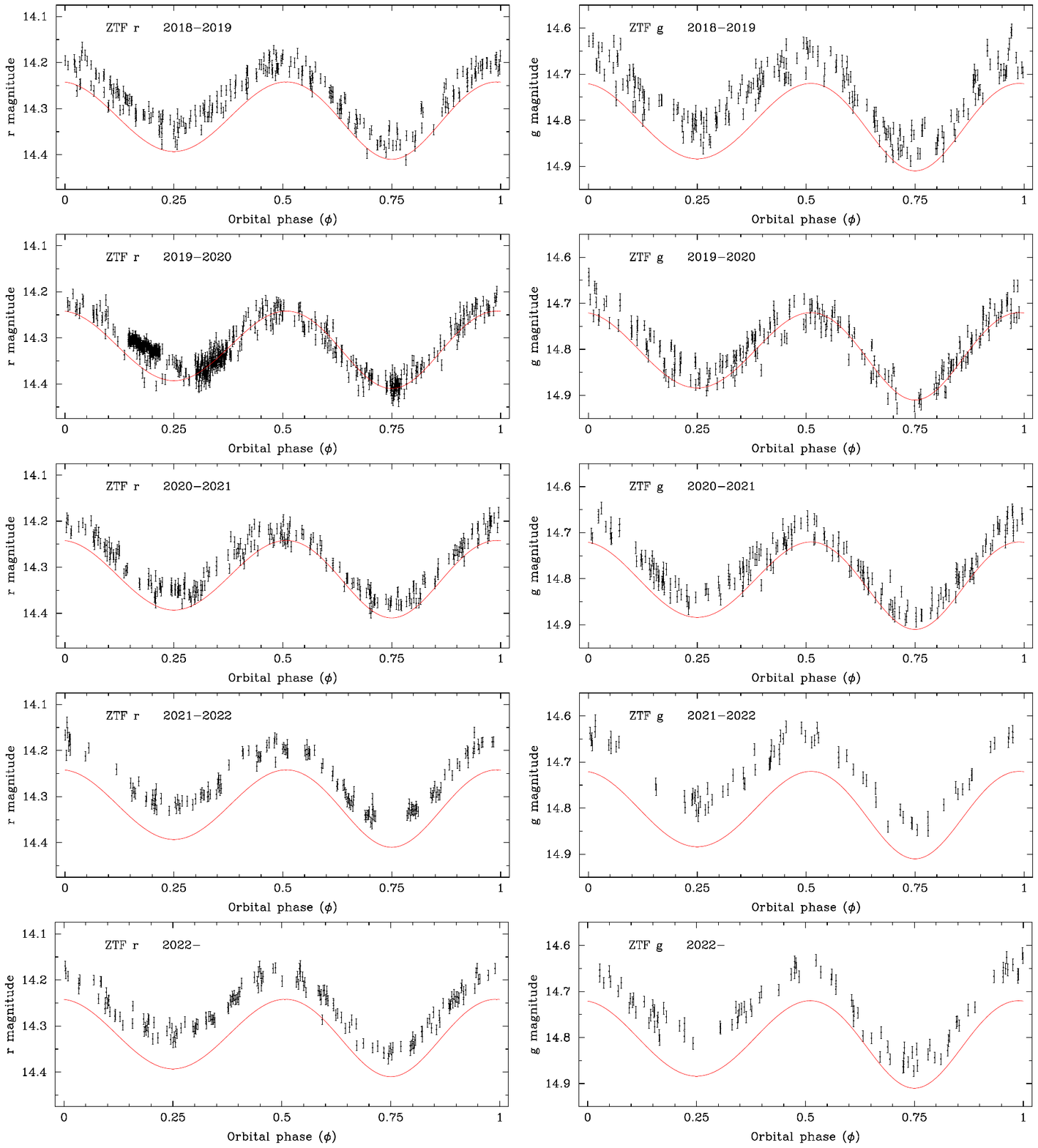}
}
\vspace{-1.in}
  \caption{ZTF observations of 3FGL J0212.1+5320 in $r$ and $g$, folded according to the orbital ephemeris calculated from the MDM data (see Section~\ref{sec:ephem}).  Each panel contains data from one annual observing season.  The red curve is the same pure ellipsoidal model used in Figure~\ref{fig:opt}, here normalized to the faintest state, which is represented by the 2019--2020 observing season.  Linear limb-darkening coefficents of 0.5 and 0.7 were used for $r$ and $g$, respectively.
}
\label{fig:ztf}
\end{figure*}

We also used the 16th data release from the ZTF (\citealt{Masci_2019,bel19}), which provided 1541 good exposures in $r$ and 907 in $g$ over the years 2018--2022.  Figure~\ref{fig:ztf} shows the ZTF points  grouped into the five observing seasons.  The most recent season (2022--2023) is not yet complete.

\subsection{Light Curve Variability}
\label{sect:optical_variability}

Dominated by ellipsoidal modulation, the light curves of 3FGL J0212.1+5320 also reveal significant variability on time scales of months.  In each panel of Figure~\ref{fig:opt}, data exhibiting consistent states are grouped together. While a detailed heating model was applied by \citet{sha17} to the original light curves of \citet{Linares_2017}, it is now apparent that different results would necessarily be obtained at other epochs.  The now familiar variability of redback companion light curves \citep{bel16,van16,Li_2014,cho18,Clark_2021,hal22} can be attributed to one or more causes: changing Roche-lobe filling factor, starspots, variable photospheric heating by the intrabinary shock, and/or heating by the magnetically ducted pulsar wind.

Although detailed modeling of these light curves is beyond the scope of this work, it is meaningful to use the basic parameters of the system that were fitted by \citet{sha17}, including mass ratio, inclination angle, and Roche-lobe filling factor, but neglecting pulsar heating and hot or cold spots, to examine how our data would compare to a model of pure ellipsoidal modulation.  For this purpose we used the fitted orbital parameters from the hot-spot model of \citet{sha17}, $q=M_2/M_1=0.28,\ i=69^{\circ}$, and volume filling factor $f=0.76$, to produce a purely ellipsoidal curve using the analytic approximations of \citet{mor93} and \citet{gom21}.  For compatibility with the analytic model, we used a linear limb-darkening coefficient $u=0.6$ appropriate for the F6 spectral type \citep{cla11}.  The red curve in Figure~\ref{fig:opt} represents this model.  The data in Figure~\ref{fig:opt}g conforms to the model very well, so we normalized the model there.  At this epoch the star was in its faintest state, with a minimum of the light curve at $V=14.60$.

Notably, the magnitudes from \citet{Linares_2017} to which \citet{sha17} fit their model also correspond closely to our data in Figure~\ref{fig:opt}g.  Their light-curve minima at $r'=14.38$ and $g'=14.97$ transform to $V=14.62$, very close to our $V=14.60$.  Data in the other panels show varying degrees of departure from the model, mostly lying above it (by up to 0.1~mag), and including deviations from pure ellipsoidal effects such as those mentioned above and reviewed in \citet{hal22}. Note that the Roche-lobe filling factor, modelled as 0.76 in volume or 0.91 in radius, could increase without leading to overfilling and mass transfer.

The ZTF light curves in Figure~\ref{fig:ztf} reinforce the evidence for long-term variability that is sparsely sampled by the MDM data.  The brightness appeared to reach a minimum in 2019--2020, just as in the MDM data, while the star was brighter in 2017--2018, and was again near maximum in 2021--2022.  In addition, within individual observing seasons, several of the ZTF light curves show excess variance around the mean ellipsoidal shape, which is indicative of variability on shorter timescales, again confirming such evidence from the MDM runs.

\subsection{New Orbital Ephemeris\label{sec:ephem}}

We used the MDM data to obtain a photometric orbital ephemeris that extends and refines the original ones from \citet{Li_2016} and \citet{Linares_2017}.  The 21~hr orbit does not allow complete phase coverage during a typical observing run.  This, plus the obvious long-term variability of the light curve, is not conducive to template fitting as a method of timing.  Instead we make use of those segments that include a well covered extremum of the light curve, either maximum or minimum, to use as fiducial markers of orbital phase.  Phase 0 is the presumed ascending node of the pulsar, corresponding to a maximum of the light curve due to the projection of the tidally distorted companion star.  The first minimum, at phase 0.25, is the presumed inferior conjunction of the companion, and so on.

We found a total of 26 useful timings of the extrema between 2017 and 2023, and supplemented these with two useful timings in $r$ and $i$ filters during 2015 December from \citet{Linares_2017}.  Fitting these 28 timings to a constant period leaves residuals of $<0.01$ cycles, with only small systematic deviations from ellipsoidal symmetry. The fitted orbital period is $P_b = 0.8695759(11)$ day, with $T_0 = 58,487.45679(86)$ BMJD as the epoch of phase 0. The span of the observations is MJD 57,378--59,971. 

Random variability of the light curve shape may account for the bulk of the error bars on the ephemeris parameters. Also, several panels in Figure \ref{fig:opt} show timing variations of the maximum and minimum of up to $\approx0.03$ cycles.  This suggests that $T_0$ is not as precise as its fitted formal uncertainty.  Since $T_0$ is a photometric quantity and not a kinematic measurement, it can be expected to deviate from the true time of ascending node of the pulsar by an unknown systematic error much larger than its statistical uncertainty.  See Section~\ref{sect:orbital_params} for further discussion of this.  

We also fitted for the orbital period in the ZTF data using a Lomb-Scargle periodogram. The value of the first harmonic, which contains more power than the fundamental, was adopted.  The resulting $P_b$ is 0.869566(7) day and 0.869575(8) day for the $r$ and $g$ points, respectively. These are both consistent with the contemporaneous MDM value, but not as precise.  Therefore, we simply folded the ZTF data in Figure~\ref{fig:ztf} on the MDM ephemeris.

\section{X-ray Observations}
\label{sect:xray}
The field around 3FGL J0212.1+5320 has been covered by multiple X-ray telescopes including NuSTAR, Swift, and Chandra. All X-ray observations of 3FGL J0212.1+5320 presented here are summarized in Table~\ref{tab:xraylog} of the Appendix. Swift UVOT magnitudes are listed in Table~\ref{tab:uvotlog}.

\subsection{NuSTAR\label{sect:nustar}}

The 3FGL J0212.1+5320 system was observed with the Nuclear Spectroscopic Telescope Array (NuSTAR; \citealt{Harrison2013}) on 2020 October 5--7 (ObsID 30601011) in a 81.7 ks deadtime-corrected on-source exposure. The total elapsed time of the observation was 168.4 ks, which covers 2.2 consecutive orbital cycles of the 3FGL J0212.1+5320 binary. The event data were processed using the \texttt{nupipeline} script in NuSTARDAS and the images, spectra, and barycentered and background-subtracted light curves were generated using the \texttt{nuproducts} task. 3FGL J0212.1+5320 is the only hard X-ray source within the NuSTAR field of view and is detected above the background level up to $\sim$30 keV. For both the spectroscopic and time-variability analyses, the source counts were extracted from a circular region of radius $60^{\prime \prime}$ (which encircles $\sim$80\% of the total energy of the point spread function) centered on the Gaia optical position.

\begin{figure}[t]
\centerline{
\includegraphics[angle=0.,width=\linewidth]{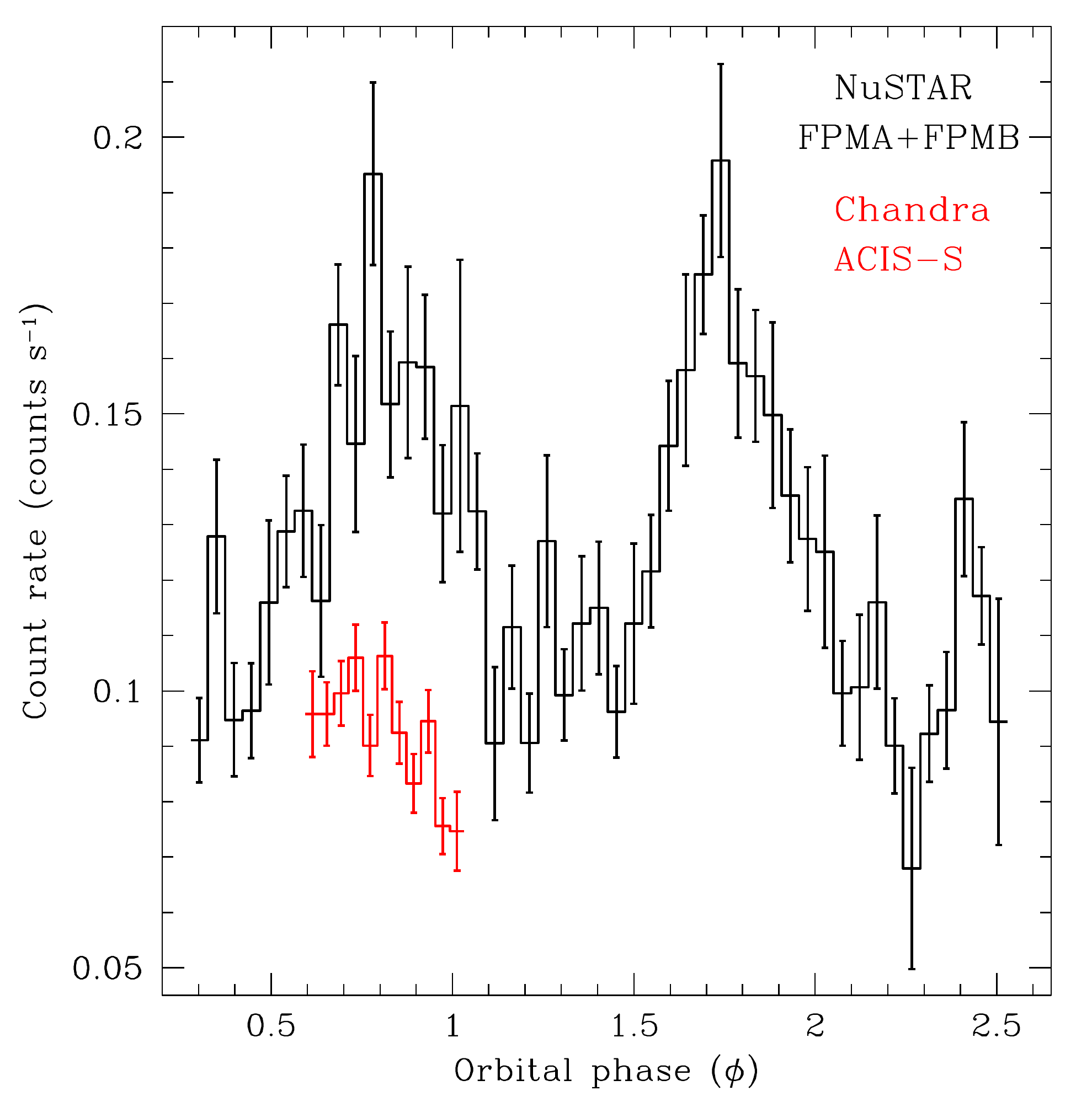}}
  \caption{NuSTAR FPMA+FPMB 3–-79\,keV background-subtracted light curve of 3FGL J0212.1+5320 binned at a resolution of 3600\,s (black) and Chandra ACIS-S 0.3-8\,keV light curve binned at 3000\,s (red). The NuSTAR count rates have been corrected for fractional exposures in each bin caused by Earth occultation. 
}
\label{fig:nustarlc}
\end{figure}

The background-subtracted NuSTAR light curve, with events from Focal Plane Modules A and B (FPMA and FPMB) combined, in the 3--79 keV band as a function of binary orbital phase binned in 5000\,s intervals is shown in Figure~\ref{fig:nustarlc}. The count rate in each bin has been corrected for the fractional exposure caused by gaps in the data due to occultation by the Earth as the telescope moves in its orbit.  Large-amplitude variability, with a factor of $\sim$2 range in flux, is clearly seen from 3FGL J0212.1+5320 and appears to repeat in the two full orbital cycles covered. The variability pattern, with a trough and peak at phases $\phi\approx0.25$ and  $\phi\approx0.75$, respectively, appears to follow the typical X-ray behavior of redbacks \citep[see, e.g,][]{Bogdanov_2011,Bogdanov_2014a,Bogdanov_2014,Gentile2014,AlNoori2018,Bogdanov2021}, which is interpreted as being due to the changing view of the intrabinary shock. This intrabinary shock forms at the interface between the relativistic pulsar wind and the outflow from the secondary star \citep{Arons_1993}. For reference, we also show a light curve based on the archival 30\,ks Chandra ACIS-S 0.3--8 keV exposure of the field around 3FGL J0212.1+5320 (ObsId 14814\footnote{Available at \url{https://doi.org/10.25574/14814}.}) obtained on 2013 August 22 (originally presented in \citealt{Li_2016} and \citealt{Linares_2017}). The Chandra light curve covers $\approx40$\% of the binary orbit around the X-ray maximum at $\phi\approx0.75$, with count rates as a function of orbital phase generally consistent with those seen in the NuSTAR data. 

\begin{figure}[t]
\centerline{
\includegraphics[angle=0.,width=\linewidth]{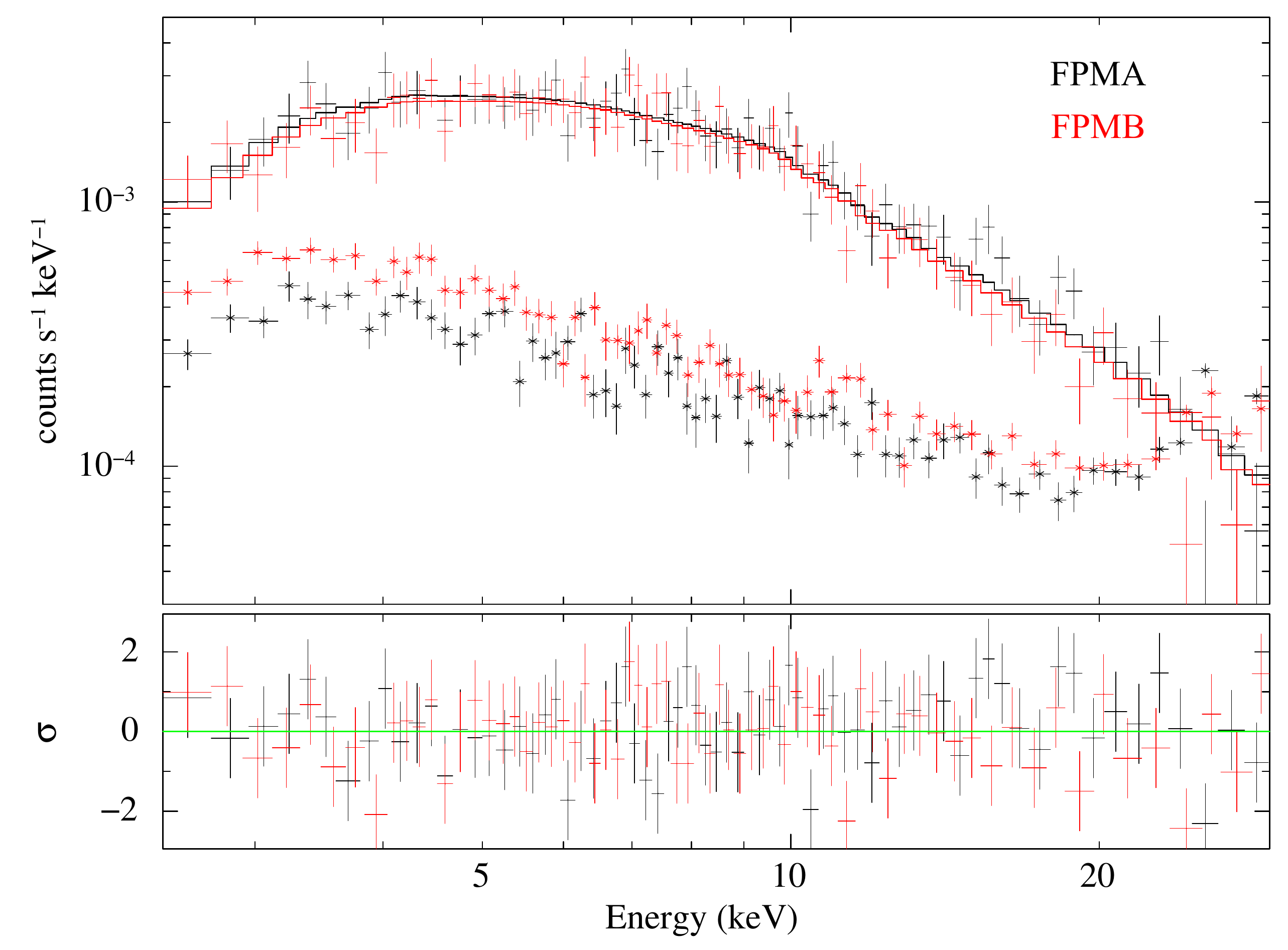}}
  \caption{Time averaged NuSTAR FPMA (black) and FPMB (red) spectra of 3FGL J0212.1+5320 fit with an absorbed power-law model. The stars show the background spectra for each detector. The bottom panel shows the best fit residuals expressed in terms of standard deviation with error bars of size 1$\sigma$. See Section \ref{sect:nustar} for best fit parameters.
}
\label{fig:nustarspec}
\end{figure}

The NuSTAR FPMA and FPMB time-averaged spectra (Figure~\ref{fig:nustarspec}) are well fit ($\chi^2=109.47$ for 136 degrees of freedom) by an absorbed power law assuming the \texttt{tbabs} model of interstellar absorption and chemical abundances from \citet{Wilms_2000}. There is no evidence for a spectral turnover at higher energies in the best fit residuals. The best fit parameter values we obtain are $N_{\rm H}<1.7\times 10^{22}$  cm$^{-2}$ for the hydrogen column density along the line of sight, $\Gamma=1.35_{-0.06}^{+0.08}$ for the spectral photon index, and a $F_X=(5.3\pm0.4)\times 10^{-12}$ erg cm$^{-2}$ s$^{-1}$ (3--79\,keV) for the unabsorbed flux, with the upper limit and all uncertainties quoted at the 90\% confidence level for one interesting parameter. For $N_{\rm H}$ fixed at the best fit value of $1.4\times10^{21}$ cm$^{-2}$ from the fit to the Chandra data\footnote{As the archival Chandra ACIS observation only has partial orbital coverage ($\approx$40\%) joint spectroscopy with NuSTAR is not feasible.} reported by \citet{Linares_2017}, we obtain comparable results with $\Gamma=1.35\pm0.06$ and $F_X=(5.3\pm0.4)\times 10^{-12}$ erg cm$^{-2}$ s$^{-1}$.  For a distance of 1.16 kpc, this corresponds to a 3--79\,keV luminosity of $L_X=8.6\times10^{32}$ erg s$^{-1}$.

\subsection{Swift XRT}
\label{sect:swiftxrt}
The Neil Gehrels Swift Observatory has targeted 3FGL J0212.1+5320 in 15 snapshot observations. We used HEASoft version 6.28 to process all XRT observations taken in photon counting mode, and produced background-subtracted light curves in the 0.3--10 keV band using the clean Level 2 event files.  We used an aperture radius of of 25 pixels (59$\arcsec$) for the source and 50 pixels (118$\arcsec$) for the background region. Within the good time intervals (GTIs), we found the correction factors for bad columns in each 10~s time interval, and applied them to 60~s time bins of the light curve.
Using the 60~s bins, we subtracted the background rate from the corrected source rate. The mean net count rate per observation is outlined in Table \ref{tab:swiftxrt_tab} and shown in Figure \ref{fig:J0212_swift_fermi}. 

To obtain sufficient statistics for spectral fitting, we group all 15 observations. We combined the exposure maps using {\fontfamily{qcr}\selectfont ximage}, and the clean Level 2 files, and extracted a spectrum from the final image using the extraction regions used in our light curves. We generated the ancillary response file using the tool {\fontfamily{qcr}\selectfont xrtmkarf}, which corrects for hot columns and bad pixels, and applied the point-spread-function correction.  We used {\fontfamily{qcr}\selectfont grppha} to bin the counts in energy, ignoring energies $<0.3$~keV. We used {\fontfamily{qcr}\selectfont group min 20} to ensure that there are at least 20 counts per bin in order to employ $\chi^{2}$ statistics. 

Using XSPEC, we fit an absorbed power-law model and integrated the flux over a 0.3--10 keV range.  
Using the same $N_{\rm H}$ of $1.4\times10^{21}$ cm$^{-2}$ \citep{Linares_2017} as in Section \ref{sect:nustar}, we obtain $\Gamma=1.18\pm0.16$ and $F_X=(1.07\pm0.15)\times 10^{-12}$ erg cm$^{-2}$ s$^{-1}$, which corresponds to a luminosity of $L_X=1.72\times10^{32}$ erg s$^{-1}$. Extrapolating this to the broader high-energy coverage of NuSTAR, we find the fluxes are consistent with each other.

\subsection{Swift UVOT}
\label{sect:swiftuvot}
Along with the XRT, 3FGL J0212.1+5320 was also observed with the UV/Optical Telescope (UVOT) with four of its six filters ($U$, $\mathit{UVW1}$, $\mathit{UVM2}$, and $\mathit{UVW2}$). The photometry was extracted using the \verb|uvotsource| command in FTOOLS. The net flux was obtained from an aperture of $5\arcsec$ radius for the source and $20\arcsec$ for the background region.  Magnitudes are in the AB system \citep{Breeveld_2011}. Table~\ref{tab:uvotlog} shows a log of the observations, including the magnitude for each image in a single filter for each observation.

\begin{figure}
\includegraphics[angle=0.,width=0.95\linewidth]{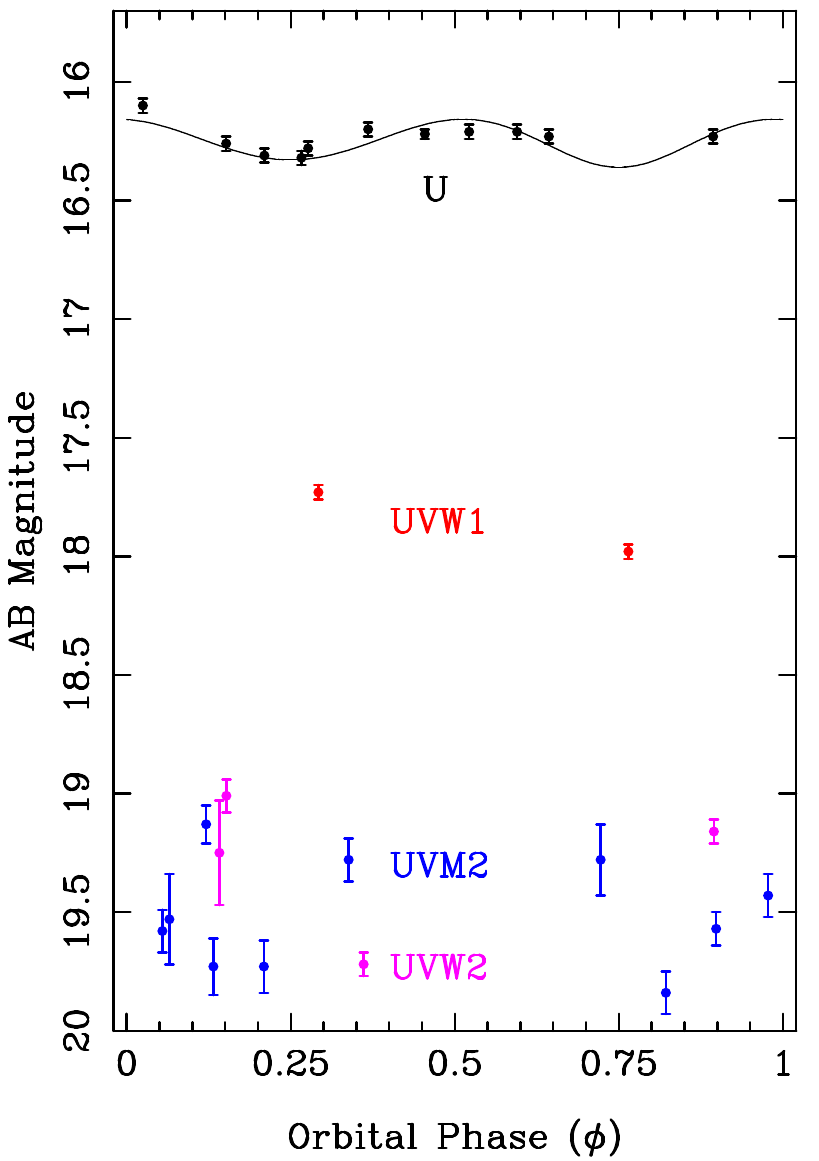}
    \caption{Swift UVOT magnitudes from Table~\ref{tab:uvotlog}, plotted as a function of orbital phase using the optical ephemeris of Table~\ref{tab:pulsar_params}.  The ellipsoidal model is superposed on the $U$ points.
}
\label{fig:uvot}
\end{figure}

In order to interpret the Swift UVOT measurements relative to the ground-based optical, we have to take into account the expected magnitudes of the F6 type companion star \citep{Linares_2017} as well as the significant extinction at this low Galactic latitude of $7^{\circ}$.  The most precise measurements are in the $U$ filter; we examined these as a function of orbital phase, finding that their variation by $\approx0.28$~mag is in accord with the phasing of ellipsoidal modulation (see Figure~\ref{fig:uvot}), although with additional scatter arising from long-term variability as is observed at visible wavelengths.

For quantitative comparison with our $V$-band measurements in Figure~\ref{fig:opt}, we convert from the AB system to the Vega system using $U_{\rm AB}-U_{\rm Vega}=+1.02$. 
\citet{gre19} evaluate the extinction as $E(g-r)=0.12\pm0.02$~mag at $d=1.15$~kpc, which corresponds to $A_V=0.41$~mag and $E(U-V)=0.24$~mag.  With these conversions and corrections applied, the intrinsic color $U_0-V_0=(U-V)-E(U-B)$ ranges from 0.41--0.49, which is consistent with the expected $U_0-V_0\approx0.47$ for an F6 main sequence star.  Therefore, the $U$-band magnitude and its variation are consistent with photospheric emission.

The $\mathit{UVW1}$, $\mathit{UVM2}$, and $\mathit{UVW2}$ measurements are more difficult to interpret either because they are sparse, their error bars are larger, and especially because extinction corrections are highly uncertain in these broad-band filters when the extinction is large (see Table~4 of \citealt{sie14}). The UV filter points are more variable than the $U$-band, but in Figure~\ref{fig:uvot} they have no clear trend with orbital phase, e.g., like the X-ray light curve.

\section{Discussion\label{sect:discussion}}

\subsection{Orbital Parameters\label{sect:orbital_params}} 
In the absence of more extensive orbital coverage of the pulsar, the binary parameters modelled by \citet{sha17} can be used to check for compatibility with the pulsar timing results of Table~\ref{tab:radio_PRESTO_params} by predicting the apparent spin period derivative $\dot P_{\rm s}$ at any orbital phase due to the acceleration.  For a circular orbit in the non-relativistic limit,
\begin{equation}
P_s = P_{s,0}\left[1 + \frac{K_2\,q}{c}\,{\rm cos}(2\pi\phi)\right],
\end{equation}
\begin{equation}
    \dot P_{\rm s} = -\frac{2\pi K_2\,q}{c}\,\frac{P_{s,0}}{P_b}\,{\rm sin}(2\pi\phi),
\end{equation}
where $P_{s,0}$ is the intrinsic spin period, $K_2 = 216.5^{+5.8}_{-5.7}$ km~s$^{-1}$ is the radial velocity amplitude of the companion, and $q = 0.28^{+0.08}_{-0.09}$ is the mass ratio. We choose to apply these to Obs 05 with its longer detection, using $\phi=0.698$, its optically determined starting phase (Table~\ref{tab:radio_obs}). The prediction is then $\dot P_{\rm s} = (3.39\pm1.09)\times10^{-11}$ s~s$^{-1}$.  Its accuracy is limited mainly by the $\approx30\%$ uncertainty in the mass ratio $q$, as well as by additional model-dependent uncertainty in $q$ when choosing authors' favored hot-spot model and neglecting their (dark) starspot model. The PRESTO measured $\dot P_{\rm s}$ of Obs 05 is $(2.987\pm 0.011)\times 10^{-11}$ s~s$^{-1}$ (Table~\ref{tab:radio_PRESTO_params}), which is consistent with the predicted value within its uncertainty, thus supporting the identification of the pulsar with the optical binary. 

Since $\dot P_{\rm s}$ is measured more precisely than $q$ is modelled from the optical, we now invert Equation (3) and solve for $q$ more accurately and precisely.  Since $K_2$ is virtually model-independent (to within $\approx 1$ km~s$^{-1}$, \citealt{sha17}), the resulting model-independent value of $q$ is $0.2469\pm0.0067$.  In addition we posit a systematic uncertainty of 0.03 on the optically determined orbital phase $\phi$ (see below), which would correspond to an error of 0.010 on $q$.  The model-independent $P_{s,0}$ is 2.1147012(46)~ms, but 2.1147(1)~ms with error dominated by orbital phase uncertainty.  Finally, the projected semi-major axis of the pulsar's orbit is
\begin{equation}
\frac{a_1\,{\rm sin}\,i}{c}=\frac{P_{b}\,K_2\,q}{2\pi c}=\frac{-\dot P_s\,P_b^{2}}{4\pi^2\,P_{s,0}\,{\rm sin}(2\pi\phi)}=2.132\pm0.008\ {\rm \text{lt-s}}.
\end{equation}
A 0.03 uncertainty on $\phi$ would also add a dominant systematic error of 0.09~lt-s on $a_1\,{\rm sin}\,i/c$.
Together with the well-measured $P_{\rm s,0}$ and $P_b$, the bound on $a_1\,{\rm sin}\,i$ will be valuable in searching for $\gamma$-ray pulsations.  

These orbital parameters imply a minimum companion mass of $M_2>0.35\,M_{\odot}$ for an assumed typical $M_1=1.4\,M_{\odot}$ primary.  If the inclination angle $i=69^{\circ}$ from the hot-spot model of \citet{sha17} is correct, then the masses become $M_1=1.75,M_{\odot}$ and $M_2=0.43\,M_{\odot}$.  If $i=64.\!^{\circ}5$ as in their dark-spot model, then $M_1=1.93\,M_{\odot}$ and $M_2=0.48\,M_{\odot}$. 

Finally, we fit the apparent spin parameters measured from both radio detections using PRESTO's {\fontfamily{qcr}\selectfont fitorb}, which uses the \citet{BTmodel_1976} binary model, to calculate preliminary parameters of the pulsar's orbit. While pulsar timing should be used to further refine these parameters, we do not yet have enough observations to obtain such a solution.  Here, our goal is to simply test for consistency with the optically-derived ephemeris. 

We allowed for eccentricity in the fit and found $e=0$, consistent with the near-perfect circular orbits of redbacks. We find an intrinsic spin period $P_{s,0}= 2.11461962(94)$, an orbital period $P_{b} = 0.869573849(10)$ days, a projected semi-major axis $(a_1/c)\,{\rm sin}\,i = 2.02169 \pm 0.00015$~lt-s, and an epoch of ascending node $T_{\rm asc} = 59757.87757(36)$ BMJD (after using the optical ephemeris extrapolated to midpoint of the two radio detections to make an initial guess of $T_{\rm asc}=59757.90718$). 
While $P_{b}$ is within $2\sigma$ of the optical photometric value and error listed in Table~\ref{tab:pulsar_params}, $T_{\rm asc}$ comes 0.034 cycles later than the prediction from the optical $T_0$, a difference that we verified by folding the optical data on the radio parameters.  This phase discrepancy can also account for the 0.11 lt-s difference in the projected semi-major axis compared with that derived above using an optically predicted phase, as well as the 0.00008 ms difference in the intrinsic spin period. 

Recall that systematic error likely dominates the uncertainty of the optical $T_0$ because it is not a kinematic parameter, but a photometric property in the context of an idealized ellipsoidal model.   With regard to the radio, the
two widely spaced detections obtained so far do not permit a phase-connected timing solution, nor provide many pulse times of arrival (TOAs) with which to carry out a long-term timing analysis and obtain precise parameters. As a result, the intrinsic spin-down rate and spin-down power of the pulsar are not yet measured. We expect that denser radio timing will pin down all of these parameters, including the time of ascending node.

\subsection{Eclipses and Flux Variability\label{sect:eclipses}}

Despite observing at inferior conjunction of the pulsar, the non-detections (Obs 01, 02, and 04) and variations in pulsar signal intensity across both time and frequency during the two detections is not unusual in redbacks as seen in PSR J1048+2339 \citep{Deneva_2016}. This suggests a possible gas cloud or an extended companion wind surrounding the pulsar \citep{cra13}, which could be related to the long-term optical flux variability we see (Section \ref{sect:optical_variability}). Eclipses covering a large fraction of the orbit are typical for redbacks ($\sim$ 50\%), especially for tMSPs \citep{Archibald_2009,Roy_2014}. This is due to redbacks having short orbital periods and large companions, as opposed to black widows which have been observed to have shorter eclipses ($\sim0.1-0.2\, P_{b}$: \citealt{Polzin_2018, pol20, Crowter_2020, den21, Ray_2022}). 

Both of our detections (Obs 03 and Obs 05) occur just before inferior conjunction of the pulsar ($\phi$ = 0.75). Although we do not assume that the detections are dependent only on orbital phase, a Smoothed Particle Hydrodynamics (SPH) simulation of the redback PSR B1744$-$24A showed the strong companion wind flowing behind and lagging the pulsar due to the combined effects of gravity, gas pressure, and the Coriolis force \citep{tav91,tav93}. This kind of asymmetric wind geometry would cause $\phi$ = 0.75--0.25 to be eclipsed, while $\phi$ = 0.25--0.75 would be visible, which is what we have observed so far (Table \ref{tab:radio_obs}). However, the \citet{tav95} simulations show pulsar enshrouding can occur for orbital periods $\lesssim 10$~hr -- a threshold that some redbacks, including this one, exceed.

In addition to absorption due to intrabinary material, another mechanism for the nondetections could be due to the interstellar medium (ISM) along the line of sight contributing to its dimming/brightening \citep{Lorimer_2004} since interstellar scintillation is stronger at lower DMs and has no preferred orbital phase (\citealt{Cordes_Weisberg_1985, Camilo_2000, Damico_2001_GCs}). PSR J1048+2339, for example, revealed the effects of scintillation after incurring different flux densities between observations at three different observing frequencies \citep{Deneva_2016}. The bright signal confined to the top of our frequency band near $\sim$ 1.8 GHz in Obs 05 (Figure \ref{fig:radio_phase_vs_subband}) and the narrow-band scintles suggests that it could be caused by scintillation since it does not occur in Obs 03, it covers a very narrow range of frequencies, and the pulsar is much dimmer in the rest of the band. 

Unlike the DM delays observed preceding black widow eclipses (PSR J1544+4937 \citealt{Bhattacharyya_2013} and PSR J1810+1744 \citealt{Polzin_2018}), PSR J1227$-$4853 \citep{kud20} and PSR 1744$-$24A \citep{nic92} have been shown to exhibit occasional delays outside of its eclipse--associated phases. \citet{nic92} observed PSR B1744$-$24A in Terzan 5 at several frequencies, and noted long-term disappearances at all frequencies, with evidence that they could be episodic in nature rather than random. PSR J1740$-$5340 in the globular cluster NGC 6397 exhibits eclipses around superior conjunction of the pulsar for 40\% of its orbit, as well as random DM variations over a wide range of orbital phases (\citealt{dam01}), while the newly found PSR J1740$-$5340b exhibits eclipses for 50\% of its orbit \citep{Zhang_2022}. PSR J2039$-$5617 \citep{cor21}, PSR J1048+2339 \citep{Deneva_2016}, and J1417$-$4402 \citep{Camilo_2016} also undergo eclipses for roughly half their orbit. PSR J1723$-$2837 was also found to exhibit flux variability and have non-detections at phases away from superior conjunction of the pulsar \citep{cra13}. \citet{kud20} conducted pulsed and continuum flux densities of PSR J1227$-$4853 and suggested cyclotron absorption by the magnetic field of the companion as a potential cause for eclipses and occasional fading, building on the original mechanism suggestion by \citet{Thompson_1994} for the BW PSR B1957+20. Based on our observations and the fading behavior seen in Figure \ref{fig:radio_phase_vs_time} , we suspect PSR J0212+5321 to also be similarly eclipsed. 

The timing anomaly observed in Obs 05 and described in Section \ref{sect:timing_anomaly} is not uncommon in eclipsing MSPs. Such kinks in timing have also been observed in 3FGL J1417.5$-$4402 (PSR J1417$-$4402; \citealt{Camilo_2016}). They could be caused by mass transfer through the L2 point during a previously unobserved accretion-powered state \citep{kud20}. Removing the kink does not affect our observed parameters, and, as a result, does not affect our derived orbital parameters. This is because the pulses during the kink do not contribute significantly to the peak of the summed pulse profile. 

\section{Conclusions\label{sect:conclusion}}
We have discovered the suspected rotation-powered MSP in the 3FGL J0212.1+5320 gamma-ray system with the Robert C. Byrd Green Bank Telescope using the L-band receiver at a center frequency of 1501 MHz utilizing the Breakthrough Listen digital receiver. PSR J0212+5321 is a $P=2.11$ ms radio pulsar with a DM of 25.7 pc\,cm$^{-3}$. PSR J0212+5321 joins three other recently discovered redbacks: PSR J0838--2827, J0955--3949 and J2333--5526, found by the TRAPUM collaboration \footnote{Discoveries listed at \url{http://www.trapum.org/discoveries/}.}(\citealt{Clark_2023_TRAPUM}). PSR J0212+5321's detection makes it one of the longest binary period eclipsing MSPs known. 

The acceleration of PSR J0212+5321 inferred from the apparent change of the spin period over time is consistent with being due to its motion in the 20.9 hour orbit determined from optical observations. The distance to the pulsar estimated from the DM is in good agreement with the Gaia parallax-derived distance. Its pulsed radio emission appears to be intermittent on timescales of weeks/months possibly due to interstellar scintillation and variable outflows of plasma from the companion star that occasionally enshrouds the pulsar. 

The NuSTAR observations presented here provide the first complete coverage of the binary orbit in X-rays, which reveals a hard non-thermal spectrum and a flux variability pattern commonly seen in redbacks. Such X-ray characteristics can be attributed to an intra-binary shock formed at the interface of the winds from the two stars. These data can provide important input for validating and improving recently developed semi-analytic models of the wind interaction in redbacks and black widows \citep{Romani16,Wadiasingh17,Kandel2019,vanderMerwe2020,Cortes22}.
The implied X-ray luminosity is one of the highest recorded among redbacks in the pulsar state and black widow MSPs in the Galactic field (\citealt{Linares_2014,Linares_2017}), followed by tMSPs PSR J1023+0038 and XSS J12270$-$4859 (\citealt{Archibald_2009, Bogdanov_2011, Roy_2014, Bogdanov_2014, Bogdanov2021}). 
The Swift XRT and UVOT coverage is fairly sparse and shows no evidence of large-amplitude X-ray variability. Judging from the lack of any significant flux changes in the long term Fermi LAT $\gamma$-ray light curve (see Figure~\ref{fig:J0212_swift_fermi} in the Appendix), there is no indication that the pulsar was accreting at the time of the radio non-detections nor at any prolonged interval in the past 14 years. However, the X-ray-to-$\gamma$-ray flux ratio of 10$\%$ shows that it is consistent with known tMSPs, which are significantly larger than those of most redbacks ($\le$ 1$\%$; \citealt{Li_2016}). This makes PSR J0212+5321 a strong tMSP candidate that may transition to a disk state in the near future and thus warrants continued close monitoring at all wavelengths.

Future observations and long-term timing of PSR J0212+5321 will allow us to measure its spin period derivative to calculate it spin-down luminosity, $\dot{E}$, which is likely one of the highest among MSPs based on its high X-ray luminosity. Although flaring behavior has not been observed in the companion like in other redbacks, the spin-down power of the pulsar is also an important factor in the energy budget of flares and any persistent heating of the companion’s photosphere \citep{hal22}. A timing solution will also allow for the detection of $\gamma$-ray pulsations, and X-ray/UV pulsations in the event that it switches to a radio-quiet disk-dominated state; thus, serving as an additional laboratory for studying the poorly understood transition mechanism of MSPs.

Conducting longer observations with a wider orbital phase coverage beyond inferior conjunction of the pulsar will help us understand the non-detections and determine the fraction of the pulsar that is eclipsed. It will also help identify any strong contrast in excess DM between these occasional delays and those occurring at the eclipse boundaries, while probing the potential effect of the magnetic field. Due to the observed pulsar brightness towards higher frequencies at L-band presented here, observations at higher frequencies, like S-band, will also help determine any frequency dependence of eclipse durations \citep{pol20}, a possible low-frequency cutoff due to absorption, and a spectral index. 

While any new available data with MDM, ZTF, Swift, NuSTAR, and Fermi will continue to be monitored and analyzed, now that the pulsar in the 3FGL J0212.1+5320 system has been detected, obtaining a radio-derived timing solution is the next step to understanding this system and will provide valuable information beyond that which optical and X-ray telescopes have so far revealed. However, given the intermittency of radio detections, it might be more feasible to obtain a $\gamma$-ray timing solution.

\begin{acknowledgments}
We thank the anonymous referee for a careful reading, several clarifications, and requests for useful additional information.

Breakthrough Listen is managed by the Breakthrough Initiatives, sponsored by the Breakthrough Prize Foundation. The Green Bank Observatory is a facility of the National Science Foundation, operated under cooperative agreement by Associated Universities, Inc. We thank the staff at Breakthrough Listen and the Green Bank Observatory for their operational support. 

We acknowledge use of computing resources from Columbia University's Shared Research Computing Facility project, which is supported by NIH Research Facility Improvement Grant 1G20RR030893-01, and associated funds from the New York State Empire State Development, Division of Science Technology and Innovation (NYSTAR) Contract C090171, both awarded April 15, 2010. 

Support for this work was provided in part by NASA NuSTAR General Observer Program grant 80NSSC21K0019 and NASA Swift Guest Investigator Program grant 80NSSC21K1380.

The MDM Observatory is operated by Dartmouth College, Columbia University, The Ohio State University, Ohio University, and the University of Michigan.  We thank Manuel Linares for sharing optical photometry data.

This work made use of observations obtained with the Samuel Oschin Telescope 48-inch and the 60-inch Telescope at the Palomar Observatory as part of the Zwicky Transient Facility project. ZTF is supported by the National Science Foundation under Grants No. AST-1440341 and AST-2034437 and a collaboration including current partners Caltech, IPAC, the Weizmann Institute for Science, the Oskar Klein Center at Stockholm University, the University of Maryland, Deutsches Elektronen-Synchrotron and Humboldt University, the TANGO Consortium of Taiwan, the University of Wisconsin at Milwaukee, Trinity College Dublin, Lawrence Livermore National Laboratories, IN2P3, University of Warwick, Ruhr University Bochum, Northwestern University and former partners the University of Washington, Los Alamos National Laboratories, and Lawrence Berkeley National Laboratories. Operations are conducted by COO, IPAC, and UW.

This work is based in part on data from the European Space Agency (ESA) mission
{\it Gaia} (\url{https://www.cosmos.esa.int/gaia}), processed by the {\it Gaia}
Data Processing and Analysis Consortium (DPAC,
\url{https://www.cosmos.esa.int/web/gaia/dpac/consortium}). Funding for the DPAC
has been provided by national institutions, in particular the institutions
participating in the {\it Gaia} Multilateral Agreement.

This research has made use of data and software provided by the High Energy Astrophysics Science Archive Research Center (HEASARC), which is a service of the Astrophysics Science Division at NASA/GSFC, the NuSTAR Data Analysis Software (NuSTARDAS) jointly developed by the ASI Space Science Data Center (SSDC, Italy) and the California Institute of Technology (Caltech, USA), data from the Chandra Data Archive, and software provided by the Chandra X-ray Center (CXC) in the application packages CIAO.

This work has relied on NASA's Astrophysics Data System (ADS) Bibliographic Services and the ArXiv.
\end{acknowledgments}

\facilities{GBT, McGraw-Hill, NuSTAR, Swift}

\software{PRESTO \citep{Ransom_2001}, DSPSR \citep{DSPSR2011}, HEASoft \citep{HEASoft2014}}

\bibliography{references}
\bibliographystyle{aasjournal}

\begin{figure*}
  \centering
  \includegraphics[scale=0.75]{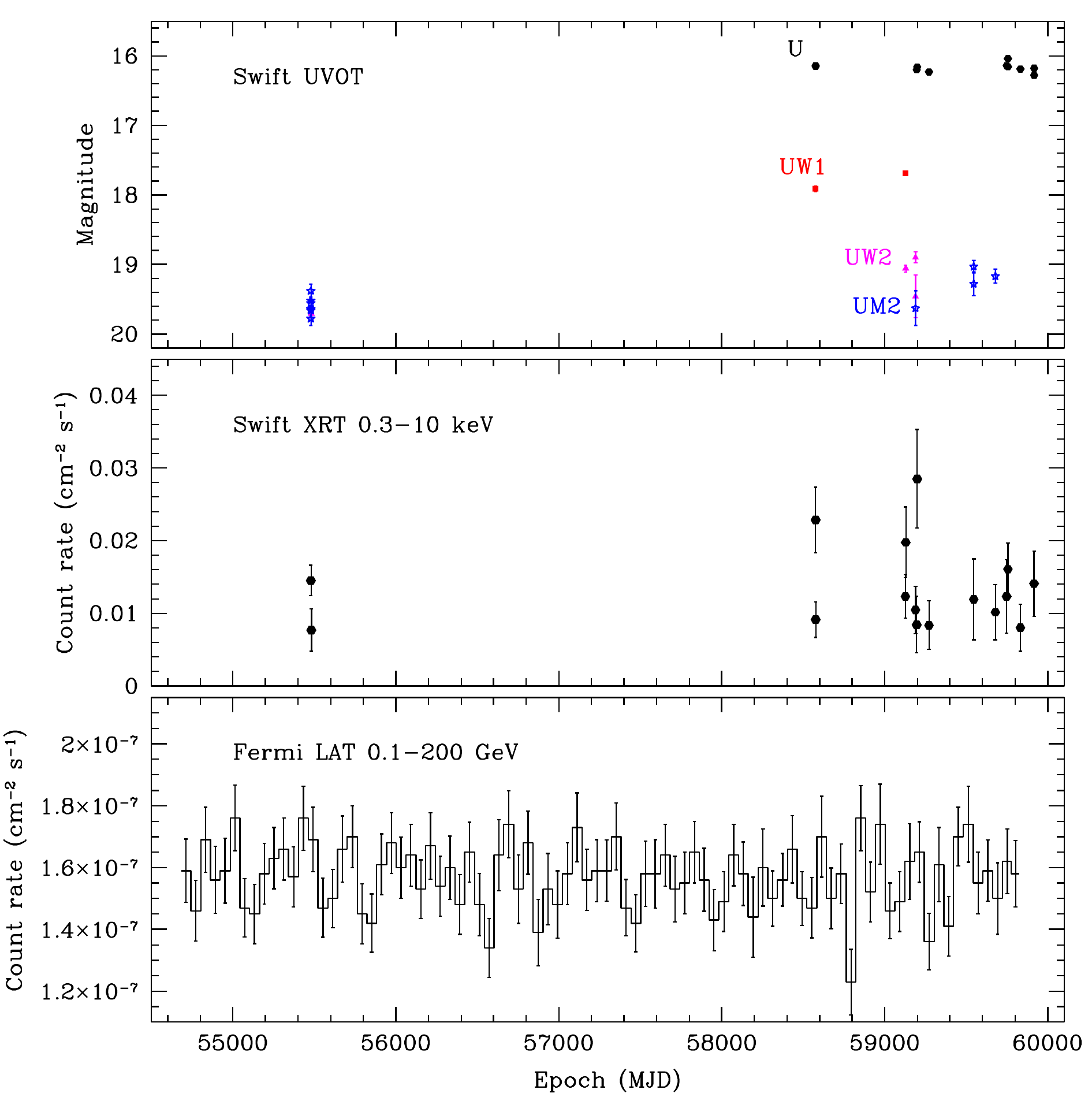}
  \caption{Long term light curves of 3FGL J0212.1+5320 from Swift UVOT and XRT (0.3--10 keV), and Fermi LAT (0.1--200 GeV) exposure-corrected aperture photometry light curve binned at 60 days. The Swift UVOT data include filters $\mathit{UVM2}$ (blue), $\mathit{UVW2}$ (magenta), $\mathit{UVW1}$ (red), and $U$ (black). }
  \label{fig:J0212_swift_fermi}
\end{figure*}

\appendix
\label{sect:appendix}
We present a log of X-ray observations of 3FGL J0212.1+5320 in Table~\ref{tab:xraylog} and the Swift UVOT exposures in Table~\ref{tab:uvotlog}. In Figure~\ref{fig:J0212_swift_fermi}, we show the long term Swift UVOT, Swift XRT (0.3--10 keV), and Fermi LAT (0.1--200 GeV) light curves. The latter was produced using the Fermitools software package from aperture photometry on data obtained from the Fermi Science Support Center\footnote{See \url{https://fermi.gsfc.nasa.gov/ssc/data/access/} for details.} and binned in 60 day intervals.

\begin{deluxetable}{cllccc}[b]
\title{test}

\tablecolumns{6} 
\tablewidth{0pt} 
\caption{Log of NuSTAR, Chandra, and Swift XRT Observations\label{tab:xraylog}}
\tablehead{
\colhead{Telescope}  & \colhead{ObsID} & \colhead{Date} & \colhead{Time} & \colhead{Exposure} & \colhead{Mean Count Rate\tablenotemark{a}} \\
& & \colhead{(UT)} & \colhead{(UTC)} & \colhead{(s)} & \colhead{(s$^{-1}$)}
}
\startdata
NuSTAR & 30601011 & 2020 Oct 5--7 & 16:46:09--15:41:09 & 81,737 & -- \\
Chandra & 14814 & 2013 Aug 22 & 04:58:55--13:55:28 & 30,119 & -- \\
Swift & 00041276001 & 2010 Oct 9 & 13:32--21:47   &  3291 &  $(1.5\pm0.2)\times10^{-2}$ \\
Swift & 00041276002 & 2010 Oct 12 & 15:20--15:41   & 1238 & $(0.77\pm0.3)\times10^{-2}$ \\
Swift & 00088697001	& 2019 Apr 2 & 22:34--22:56   &  1286  & $(2.3\pm0.5)\times10^{-2}$ \\
Swift & 00088697003	& 2019 Apr 3 & 12:55--14:36   &  1926 & $(0.91\pm0.2)\times10^{-2}$ \\
Swift & 00089046001	& 2020 Oct 5 & 17:00--17:28   &   1649 & $(1.2\pm0.3)\times10^{-2}$ \\
Swift & 00089046002	& 2020 Oct 7 & 02:27--02:55   &  1670 & $(2.0\pm0.5)\times10^{-2}$ \\
Swift & 00095748002	& 2020 Dec 6--7 & 07:50--03:09  & 821 & $(1.0\pm0.3)\times10^{-2}$  \\
Swift & 00095748004	& 2020 Dec 13 & 06:54--07:08   &  832 & $(0.84\pm0.4)\times10^{-2}$ \\
Swift & 00095748005 & 2020 Dec 16 & 06:43--07:02   &  1108 & $(2.9\pm0.7)\times10^{-2}$ \\
Swift & 00095748006 & 2021 Feb 27 & 01:10--01:26   &  960 & $(0.83\pm0.3)\times10^{-2}$ \\
Swift & 00095812001 & 2021 Nov 28 & 00:28--08:38   &  980 & $(1.2\pm0.6)\times10^{-2}$ \\
Swift & 00095812003 & 2022 Apr 9 & 21:01--21:15   &  847 & $(1.0\pm0.4)\times10^{-2}$ \\
Swift & 00096553001 & 2022 Jun 18 & 05:51--06:06   &  915 & $(1.2\pm0.5)\times10^{-2}$ \\
Swift & 00095812004 & 2022 Jun 26 & 04:25--17:32   &  1427 & $(1.6\pm0.4)\times10^{-2}$ \\
Swift & 00096553002 & 2022 Sep 10 & 22:00--22:15   &  920  & $(0.80\pm0.3)\times10^{-2}$ \\
Swift & 00096553003 & 2022 Dec 3 & 08:25--11:52 & 1121 & $(1.4\pm0.4)\times10^{-2}$ \\
\enddata
\tablenotetext{a}{Due to the Swift coverage being sparse, we calculated only the mean net count rate ($0.3-10$~keV) per observation, as described in Section \ref{sect:swiftxrt}.}

\label{tab:swiftxrt_tab}
\end{deluxetable}

\begin{deluxetable}{llccccc}
\label{tab:uvotlog}
\tablecolumns{7} 
\tablewidth{0pt} 
\tablecaption{Log of Swift UVOT Photometry}
\tablehead{
\colhead{ObsID} & \colhead{Date} & Time\tablenotemark{a} & \multicolumn4{c}{Magnitude} \\
& \colhead{(UT)} &  (MJD UTC) & \colhead{$U$} & \colhead{$\mathit{UVW1}$} & \colhead{$\mathit{UVM2}$} & \colhead{$\mathit{UVW2}$}
}
\startdata
00041276001 & 2010 Oct 9 & 55478.5683 & \nodata & \nodata & 19.84(09) & \nodata \\
& & 55478.6350 & \nodata & \nodata & 19.57(07) & \nodata \\
& & 55478.7040 & \nodata & \nodata & 19.43(09) & \nodata \\
& & 55478.7707 & \nodata & \nodata & 19.58(09) & \nodata \\
& & 55478.8383 & \nodata & \nodata & 19.73(12) & \nodata \\
& & 55478.9053 & \nodata & \nodata & 19.73(11) & \nodata \\
\hline
00041276002 & 2010 Oct 12 & 55481.6463 & \nodata & \nodata & \nodata & 19.72(05) \\
\hline
00088697001 & 2019 April 2 & 58575.9479 & \nodata & 17.98(03) & \nodata & \nodata \\
\hline
00088697003 & 2019 April 3  & 58576.5478 & 16.22(02) & \nodata & \nodata & \nodata \\
& & 58576.6066 & 16.21(03) & \nodata &\nodata  & \nodata \\
\hline
00089046001 & 2020 Oct 5 & 59127.7179 & \nodata & 17.73(03) & \nodata & \nodata  \\
\hline
00089046002 & 2020 Oct 7 & 59129.1116 & \nodata & \nodata & \nodata & 19.16(05)  \\
\hline
00095748002 & 2020 Dec 6--7 & 59189.3264 &\nodata &\nodata & \nodata & 19.25(22)  \\
&  & 59189.3360 &\nodata &\nodata & \nodata & 19.01(07)  \\
&  & 59190.1299 &\nodata &\nodata & 19.53(19) & \nodata \\
\hline
00095748004 & 2020 Dec 13 & 59196.2921 & 16.26(03) &\nodata &\nodata &\nodata \\
\hline
00095748005 & 2020 Dec 16 & 59199.2864 & 16.21(03) & \nodata &\nodata & \nodata \\
\hline
00095748006 & 2021 Feb 27 & 59272.0539 & 16.28(03) & \nodata & \nodata & \nodata \\
\hline
00095812001 & 2021 Nov 28 & 59546.0239 & \nodata & \nodata & 19.28(09) & \nodata \\
& & 59546.3580 & \nodata & \nodata &  19.28(15) & \\
\hline
00095812003 & 2022 April 9 & 59678.8802 & \nodata & \nodata & 19.13(08) & \nodata \\
\hline
00096553001 & 2022 June 18 & 59748.2486 & 16.23(03) & \nodata & \nodata & \nodata \\
\hline
00095812004 & 2022 June 26 & 59756.1886 & 16.10(03) & \nodata & \nodata & \nodata \\
& & 59756.7268 & 16.23(03) & \nodata & \nodata & \nodata \\
\hline
00096553002 & 2022 Sep 10 & 59832.9214 & 16.32(03) & \nodata & \nodata &  \nodata\\
\hline
00096553003 & 2022 Dec 3 & 59916.3518 & 16.31(03) & \nodata & \nodata & \nodata \\
& & 59916.4890 & 16.20(03) & \nodata & \nodata & \nodata \\
\enddata
\tablenotetext{a}{Mid-time of the exposure.}
\end{deluxetable}

\end{document}